\documentclass[fleqn,usenatbib]{mnras}

\usepackage{newtxtext,newtxmath}

\usepackage[T1]{fontenc}

\DeclareRobustCommand{\VAN}[3]{#2}
\let\VANthebibliography\thebibliography
\def\thebibliography{\DeclareRobustCommand{\VAN}[3]{##3}\VANthebibliography}

\usepackage{graphicx}	
\usepackage{amsmath}	
\usepackage{fontawesome5}
\usepackage{hyperref}
\usepackage{newtxtext,newtxmath}

\makeatletter
\newcommand{\github}[1]{%
   \href{#1}{\faGithubSquare}%
}
\makeatother

\newcommand{\lenstronomy}{\textsc{lenstronomy}}
\newcommand{\gradZ}{ \nabla Z }
\newcommand{\Hii}{H\textsc{ii} }

\title[Forward-modelling galaxy metallicity distributions]{A forward-modelling approach to overcome PSF smearing and fit flexible models to the chemical structure of galaxies}

\author[B. Metha et al.]{
Benjamin Metha$^{1,2,3}$\thanks{E-mail: methab@student.unimelb.edu.au},
Simon Birrer$^{4}$,
Tommaso Treu$^{3}$,
Michele Trenti$^{1,2}$,
Xuheng Ding$^{5}$,
Xin Wang$^{6,7,8}$
\\
$^{1}$School of Physics, The University of Melbourne, VIC 3010, Australia\\
$^{2}$ARC Centre of Excellence for All Sky Astrophysics in 3 Dimensions (ASTRO 3D), Melbourne, VIC 3000, Australia\\
$^{3}$Department of Physics and Astronomy, University of California, Los Angeles, 430 Portola Plaza, Los Angeles, CA 90095, USA\\
$^{4}$Department of Physics and Astronomy, Stony Brook University, Stony Brook, NY 11794-3800, USA\\
$^{5}$Kavli Institute for the Physics and Mathematics of the Universe (Kavli IPMU, WPI), The University of Tokyo, Chiba 277-8583, Japan\\
$^{6}$School of Astronomy and Space Science, University of Chinese Academy of Sciences (UCAS), Beijing 100049, China\\
$^{7}$National Astronomical Observatories, Chinese Academy of Sciences, Beijing 100101, China\\
$^{8}$Institute for Frontiers in Astronomy and Astrophysics, Beijing Normal University, Beijing 102206, China\\
}

\date{Accepted XXX. Received YYY; in original form ZZZ}

\pubyear{2023}

\begin{document}
\label{firstpage}
\pagerange{\pageref{firstpage}--\pageref{lastpage}}
\maketitle

\begin{abstract}
\href{https://github.com/astrobenji/lenstronomy-metals-notebooks}{\faGithubSquare} 
Historically, metallicity profiles of galaxies have been modelled using a radially symmetric, two-parameter linear model, which reveals that most galaxies are more metal-rich in their central regions than their outskirts. However, this model is known to yield inaccurate results when the point-spread function (PSF) of a telescope is large. Furthermore, a radially symmetric model cannot capture asymmetric structures within a galaxy. In this work, we present an extension of the popular forward-modelling python package \lenstronomy, which allows the user to overcome both of these obstacles. We demonstrate the new features of this code base through two illustrative examples on simulated data. First, we show that through forward modelling, \lenstronomy\ is able to recover accurately the metallicity gradients of galaxies, even when the PSF is comparable to the size of a galaxy, as long as the data is observed with a sufficient number of pixels. Additionally, we demonstrate how \lenstronomy\ is able to fit irregular metallicity profiles to galaxies that are not well-described by a simple surface brightness profile. This opens up pathways for detailed investigations into the connections between morphology and chemical structure for galaxies at cosmological distances using the transformative capabilities of JWST. Our code is publicly available and open source, and can also be used to model spatial distributions of other galaxy properties that are traced by its surface brightness profile.
\end{abstract}




\section{Introduction}

Chemical enrichment of the interstellar medium is a powerful diagnostic of galaxy formation and evolution. The abundance of elements such as Carbon, Nitrogen, and Oxygen carries information about the star formation and merger history and about the accretion of pristine gas from the surrounding intergalactic medium \citep[e.g.][]{Maiolino+Mannucci19}.

Studying spatially resolved chemical abundances as a function of cosmic time has long being hailed as one of the keys to understanding the processes that govern star formation (e.g. feedback and superwinds) as well as disentangling galactic growth mechanisms such as in-situ star formation, ex-situ star formation, major and minor mergers, cold-flow accretion, and environmental effects \citep{Tremonti+04, Zahid+12, Zahid+14, Gibson+13, Jones+13, Ho+15, Belfiore+19, Franchetto+21}.

In the local Universe, much progress has been achieved in the past few years through the wholesale study of large samples of galaxies with integral field spectrography (IFS) \citep[e.g.][]{CALIFA, SAMI, MANGA}. A picture of the chemical structure of local spiral galaxies has emerged. In general, chemical abundance declines with radius, described by a so-called negative chemical gradient \citep[e.g.][]{Searle71, VilaCostas+92, Sanchez+14, Bresolin+Kennicutt15, Berg+13, Berg+20, Sanchez-Menguiano+18, Poetrodjojo+21}, which is interpreted as the result of inside-out star formation (the buildup of metals in the central regions over successive generations of star formation; e.g. \citealt{Boissier+Prantzos99, Kobayashi+11, Pilkington+12}). 

However, not all galaxies in the local Universe have metallicity profiles that follow this trend. Many studies of the abundance distributions of local galaxies have found that the metallicity does not decrease constantly throughout the entirety of a galaxy disc, showing flatter slopes in its inner regions  \citep[e.g.][]{Belley+Roy92, Zinchenko+16}, its outskirts \citep[e.g.][]{Martin+Roy95, Worthey+05, Bresolin+09, Bresolin+11, Goddard+11, Lepine+11, Marino+12, Yong+12, Sanchez+14, Marino+16}, or both \citep[e.g.][]{Rosales-Ortega+11, Sanchez+15, Sanchez-Menguiano+18}. The most extreme deviations from this inside-out model show "inverted" gradients, in which metallicity increases with radius rather than decreasing \citep{Sanchez+14, Perez-Montero+16}.

Modelling the chemical structure of galaxies at $z\gtrsim1$
would allow us to deepen our understanding of how the interstellar media (ISM) of galaxies evolve, revealing how the processes that govern star formation have been changing over time. However, there are several problems that prevent the models that have been discussed thus far from being able to provide an insightful description of the chemical profile of galaxies at cosmic noon.

First of all, at these redshifts, spatially resolved metallicity measurements for galaxies are extremely difficult to obtain. Observational challenges include the faintness of distant galaxies, the need for infra-red wavelength coverage, and the limited angular resolution compared to the typical apparent size of galaxies. Having a poor angular resolution has been shown to flatten  the inferred metallicity gradients of galaxies \citep[e.g.][]{Yuan+13, Mast+14, Acharyya+20} especially in their inner regions \citep{Belfiore+17}, which may lead to erroneous results if it is not corrected for.

Secondly, the models discussed thus far that deviate from a singly-sloped negative metallicity gradient are not a priori physically motivated -- rather, they are motivated by the fact that a simple negative linear gradient does not seem to fit the data. For this reason, many different theories have been proposed \textit{post-hoc} to explain how the modelled radial profiles may have emerged. The presence of an inverted metallicity gradient may be interpreted as being due to strong tidal forces resulting from interactions \citep{Kewley+06, Kewley+10, Rupke+10, Torrey+12}, or from  enriched gas being blown away by powerful winds driven by stupendous starbursts, while cold pristine gas is being accreted directly to the centre via cold-flows \citep{Dekel+09}. Flatter metallicity profiles in the outskirts of galaxies could be caused by a lack of sufficient material for star formation at large radii, preventing enrichment \citep[e.g.][]{Rosales-Ortega+11}; galactic fountains wherein outflowing material is re-accreted onto the galaxy at large galactocentric distances \citep{Belfiore+17}; the accretion of metal poor satellite galaxies \citep{Bird+12}; radially symmetric gas inflow \citep[e.g.][]{Portinari+Chiosi2000}; or angular momentum transport \citep[e.g.][]{Lacey+Fall85}, possibly driven by spiral arms, bars, or dynamical resonances \citep[e.g.][]{Martin+Roy95, Minchev+11}. In the inner regions, proposed explanations for flattening of the abundance profile include accumulation of gas due to the tidal influence of bars or spiral arms \citep{Rosales-Ortega+11, Sanchez+11} or at the inner Lindblad resonance \citep{Fathi+07}; but this could also simply be the effect of beam smearing in poorly-resolved data \citep{Belfiore+17}. Overall, the presence of breaks in a radial metallicity profile indicate that new processes beyond simple inside-out formation must be at play in the formation history of a galaxy, but they alone do not provide us with enough information to determine what those other processes are.

\begin{figure*}
    \centering
    \includegraphics[width=0.95\textwidth]{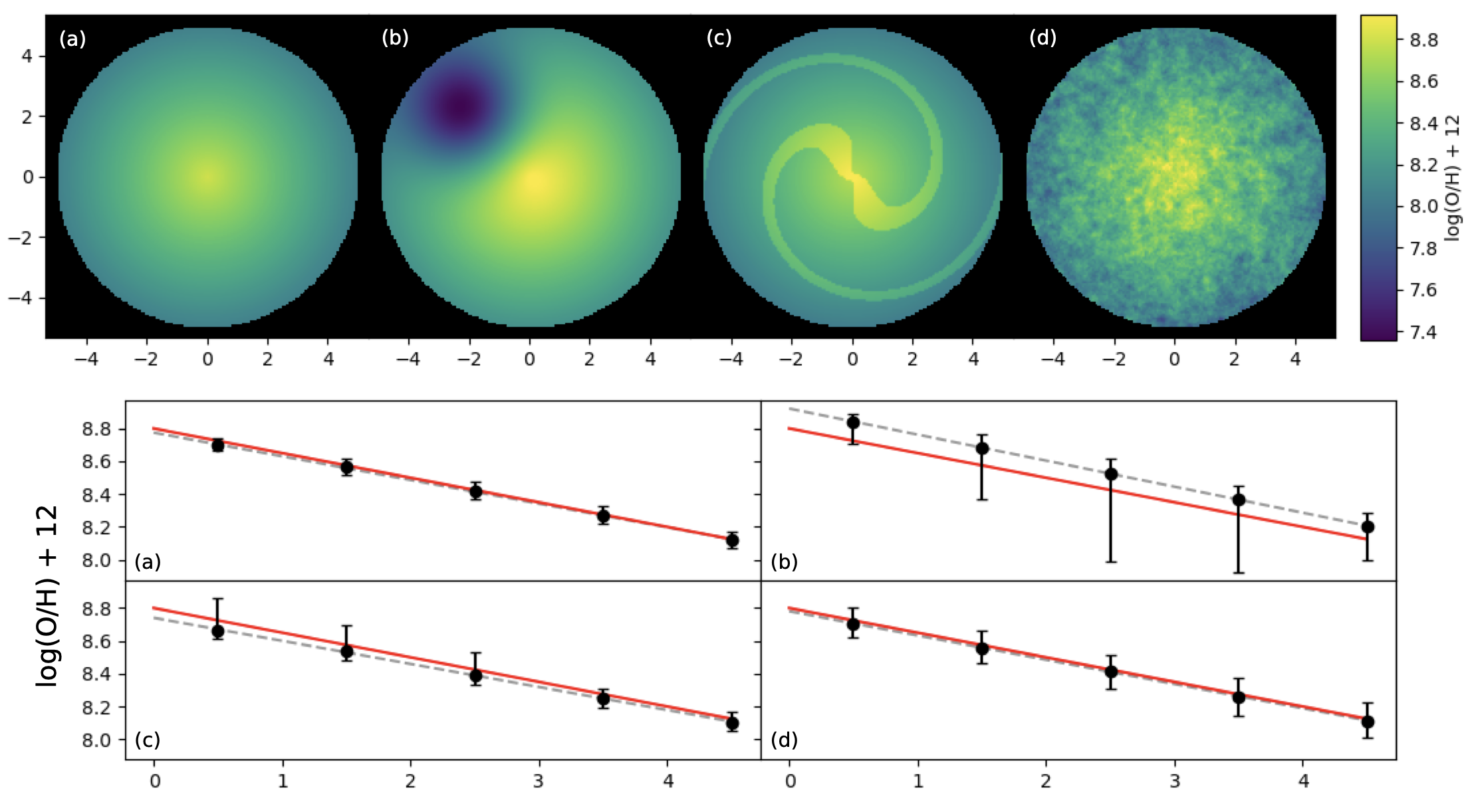}
    \caption{\emph{Top:} The four different simulated metallicity profiles, with no noise, representing four different astronomical scenarios: (a) a smooth profile with a negative metallicity gradient, representing inside-out star formation; (b) an asymmetric metallicity profile, representing a galaxy that has recently undergone a major merger with a metal-poor companion; (c) a galaxy with metal-enrichment along its spiral arms; (d) a galaxy with small-scale correlated metallicity fluctuations, representing inefficient metal mixing. \emph{Bottom:} The radial metallicity profile of these four galaxies. Red solid lines show the best fit metallicity profile from an ordinary least-squares fit using all pixels. Despite the clear differences in their 2D profiles, this is exactly the same for each galaxy. Black points with error bars show the median, 16$^\text{th}$ and 84$^\text{th}$ percentiles of the metallicity in five annuli of equal width for each galaxy. Grey dashed lines are the best-fit metallicity profile using the median metallicity in these annuli instead. For cases (b) and (c), where there are large asymmetrical structures in the metallicity profiles of the galaxies, these two commonly-used methods do not recover the same radial metallicity profiles.} 
\label{fig:same_gradZ_diff_galaxies}
\end{figure*}

Thirdly, all of the models mentioned above assume that the metallicity profile of a galaxy is radially symmetric. There are many astrophysical reasons why this is not expected to be the case at high redshifts:
\begin{enumerate}
    \item The frequency of minor and major mergers increases with redshift \citep{Conselice+Arnold09}. Mergers between galaxies of different metallicities ought to leave visible spatially correlated chemical signatures that will persist until the two galaxies are well-mixed.
    \item Star forming galaxies at cosmic noon and beyond are generally not symmetric, with the fraction of clumpy and irregular galaxies increasing with redshift \citep{Conselice14}. The circularly-symmetric metallicity gradient model was developed to describe stable disc galaxies, is probably only apt to describe galaxies with Sa/Sb/Sc morphologies, and does not naturally apply to irregular systems.
    \item Pristine gas is expected to be accreted along filaments, not in a spherically symmetric fashion \citep{Keres+05}, causing asymmetric metal poor regions inside a galaxy.
    \item Winds/outflows generally produce asymmetric regions of enriched gas \citep{Martin+02, Cameron+21}, and those gas-rich regions are not expected to fall back in a spherically symmetric way. 
    \item Spiral arms and bars are expected to produce azimuthal variations in the chemical structure of galaxies, either due to enhanced star formation along the spiral arms producing more metals \citep{Spitoni+19} or from secular processes driving radial gas flows that bring metal poor gas in towards the galaxy centre and drive enriched gas out \citep{Grand+16}.
\end{enumerate}
All of these astrophysical processes will leave observable signatures in the spatial distribution of metallicity within the galaxy that cannot be distinguished by a metallicity gradient (we demonstrate this in Figure ~\ref{fig:same_gradZ_diff_galaxies}).

However, given the observational challenges, most studies on the metallicity of the high-redshift Universe to date have focused on deriving integrated gas phase metallicities, \citep[e.g.][]{Sanders+20, Sanders+21, Sanders+23} or fitting simplified description of the chemical structure, such as a singly-sloped metallicity gradient model \citep[e.g.][]{Leethochawalit+16, Wang+22}. Galaxies with strongly "inverted" gradients that do not agree with a simple inside-out formation model have recently been discovered at high redshift \citep{Cresci+10, Queyrel+12, Swinbank+12, Stott+14, Troncoso+14, Wuyts+16, Wang+19} at a frequency that is much higher than what is generally predicted by state-of-the-art theoretical models, challenging current implementations of feedback \citep{Ma+17, Hemler+21, Tissera+22}. One reason may be because these azimuthally-asymmetric chemical features are not able to be detected by a metallicity gradient \citep{Tissera+22}. Indeed, recent simulations have shown that galaxies with very large positive or negative metallicity gradients show signs of morphological disturbances \citep{Tissera+16}.

Attempts have been made to find azimuthal variations in the abundance profiles of galaxies in the local Universe with a variety of statistics. \citet{Zinchenko+16} defined the ``global azimuthal abundance asymmetry" as the difference between the average deviations from a singly-sloped metallicity profile between two half-planes of a galaxy, and used this statistic to search for large scale asymmetries in galaxies from the CALIFA DR2 catalogue. They found that this statistic was highly degenerate with uncertainties in the central position of the galaxy used for a radial fit, and hence could not find definitive evidence of any asymmetries in the data with this method. A related approach is to split galaxies into half-planes \citep[e.g.][]{Li+13} or quadrants \citep[e.g.][]{Rosales-Ortega+11}, fit a linear profile to each sector, and search for statistically significant differences in the metallicity gradient between different regions.
However, such models implicitly predict large discontinuities at a fixed angle of azimuth, which is not physically motivated. While they could potentially identify large scale azimuthal variations in the abundance profiles of galaxies, they have limited power in explaining where such variations come from.

Many studies have searched for evidence of azimuthal variation by measuring the scatter in the metallicity around a radially-symmetric abundance profile \citep[e.g.][to name a few]{Martin+Roy95, Rosolowsky+Simon08, Bresolin+11, Croxall+15, Croxall+16, Grasha+22}. In addition to not being able to explain the origin of any metallicity variations that are found, this method is confounded by the large uncertainties in metallicity diagnostics, especially present when strong emission line calibrations are used to determine the metallicity \citep{Kewley+Ellison08}. As such, metallicity variations found with this approach are treated as upper limits to any true azimuthally varying component. \citet{Metha+21} demonstrate how uncorrelated noise in metallicity estimates can be separated from small-scale azimuthal variations using the semivariogram, a mathematical tool adopted from geostatistics. However, this geostatistical approach requires high-resolution galaxy data, with sampling on scales of a few 100 pc (Metha et al. 2024 in prep.), making them unsuitable for the majority of high-redshift galaxy data currently available. 

Another approach is to search for correlations between deviations from a radially symmetric metallicity profile and asymmetric features in a galaxy's surface brightness profile. This method has found some success in identifying asymmetric metallicity profiles in the local Universe and providing physically justified explanations for their origins. Several studies have found enrichment associated with the spiral arms of well-resolved galaxies \citep[e.g.][]{Martin+Roy95, Cedres+12, Sanchez-Menguiano+16, Sanchez-Menguiano+18, Sanchez-Menguiano+20, Ho+17, Ho+18, Vogt+17, Sakhibov+18}. For some of these studies, these variations show an enrichment on the trailing edge of the spiral arm and a dilution on the leading edge, which implies that spiral arms drive large scale radial gas flows that bring metal-poor gas into the galaxy and drive metal-rich gas out, in agreement with zoom-in galaxy simulations \citep{Grand+15, Grand+16}. Other studies show enrichment on the spiral arms, which can be attributed to enhanced star formation \citep{Spitoni+19, Molla+19}. \citet{Ho+17} note a slightly different trend wherein spiral arms appear more metal rich with a rapid drop off in the metallicity at the leading edge and a gradual decline at the trailing edge, which they explain through a cycle of self-enrichment of \Hii regions as they approach a spiral arm, followed by a rapid dilution after the density wave has been crossed. In all of these cases, morphological information from the light profiles of these galaxies is used to not only identify chemical substructures, but also to provide explanations on the origins of these features, providing a more comprehensive picture of galaxy evolution for these targets.

In this study, we present a method that enables physically-motivated models to be fit for high redshift galaxies, wherein deviations from a one-dimensional metallicity profile are correlated with asymmetrical features in a galaxy's surface brightness profile. We implement this method in \lenstronomy\footnote{\url{github.com/lenstronomy/lenstronomy/}} \citep{Birrer+Amara18, Birrer+21}, an open-source public general usage astronomy Python package that is widely used in the community. This extension of \lenstronomy\ allows two-dimensional models to be fit to a galaxy's chemical structure in a \emph{general} and \emph{rigorous} way. It is rigorous in that it accounts for the effects of pixelisation and PSF smearing using \lenstronomy's forward-modelling approach, allowing us to accurately recover metallicity gradients of distant galaxies at $z \sim 2$, whereas traditional approaches would be biased; and it is general in that the code we present is entirely flexible, allowing the user to describe the distribution of metallicity with multiple components, as required by the data. This allows for large-scale deviations from a radially-symmetric trend to be captured, and allows these deviations to be associated to different morphological components of galaxies.

We organise this paper as follows. After a brief demonstration that motivates moving beyond a radial metallicity gradient (Section \ref{ssec:phys_motivation}),
in Section~\ref{sec:newcode} we describe how we implement pixellisation and PSF smearing for light-weighted data in \lenstronomy. In Section~\ref{sec:gradient}, we demonstrate the code by applying it to a simple gradient model, in the presence of PSF smearing and pixelisation. In Section~\ref{sec:clumpy} we illustrate the ability of the code to describe clumps in metallicity. Section~\ref{sec:discussion} discusses the broader applications of the code. Section~\ref{sec:summary} provides a brief summary.

\section{Motivation}
\label{ssec:phys_motivation}

In this Section, we present a very simple illustration on how the assumption of circular symmetry can be misleading in understanding the chemical distributions of galaxies. In Figure~\ref{fig:same_gradZ_diff_galaxies}, we show 2D metallicity profiles for four different simulated galaxies. Each of these profiles are meant to represent idealised versions of a different astrophysical scenario. Panel (a) shows a galaxy with a metallicity gradient that is efficiently mixed in the azimuthal direction. Panel (b) has a large asymmetry in its metallicity profile, as may be expected from the recent accretion of a metal-poor satellite. Panel (c) shows a galaxy with a chemical enrichment of $0.2$ dex along its spiral arms, motivated by the observations of \citet{Sanchez-Menguiano+16, Ho+17}, and \citet{Vogt+17}, reflecting a possible consequence of the enhanced star formation within spiral arms. In panel (d), we show a galaxy with correlated local fluctuations in its metallicity profile on small-scales, as found by recent studies of the local Universe \citep[e.g.][]{Metha+21, Metha+22, Li+21, Li+23}, representing inefficient mixing of metals in the ISM. 

Below their 2D profiles, we plot the radial metallicity profiles that would be recovered for each of these toy models, using two different common methods from the literature. In red, we show the best-fit relation between metallicity and radius computed using an ordinary least-squares approach, using every pixel as a data point. When this method is used, these four model galaxies have identical radial metallicity profiles. Although they are clearly distinct in two dimensions, reflecting the different physical processes at play, this information is lost when a one-dimensional model is fit. 

In black, we show the median metallicity (with the 16$^\text{th}$ and 84$^\text{th}$ percentiles shown as errorbars) computed in 5 radial annuli of equal width for these galaxies. Grey dashed lines show the best-fit metallicity profiles to this azimuthally-averaged data. We find that for scenarios (b) and (c), where the metallicity profile is asymmetric on large scales, the metallicity profiles computed using all pixels and radially-averaged annuli do not agree. This disagreement is not a consequence of either method being less accurate than the other; rather, it is due to a fundamental failure of the radially-symmetric metallicity gradient model to describe the true chemical distributions of these galaxies.

From this demonstration, we can see how using a model that assumes circular symmetry could confound the interpretation of high redshift data, where metallicity maps are often clumpy and irregular \citep[e.g.][]{Cresci+10, Forster-Schreiber+18, Wang+20, Curti+20b}. For example, the accretion of a high metallicity satellite galaxy could mimic an inverted gradient, even though the formation pathways for these two scenarios are clearly distinct.

\section{Methodology}
\label{sec:newcode}

Metallicity, as we observe it, is a light-weighted quantity. The metallicity measured within an aperture is not simply the average mass of elements heavier than Helium divided by the total mass of elements contained within that region. Instead (assuming unbiased metallicity diagnostics), it is the average metallicity of the spectral energy density of all photons that fall within that aperture. 

Let $Z(\vec x)$ and $I(\vec x)$ be the intrinsic metallicity (reported in linear, not logarithmic units)\footnote{We note that these methods are not sensitive to the absolute value of the metallicity reported, and will work equally well in computing the effects of PSF smearing and blending between sources for any quantity that is linearly proportional to the metallicity. Because of this, our methods are not affected by the large systematic offsets between different metallicity calibrators that are known in the literature \citep{Kewley+Ellison08}.
} and surface profile of a galaxy, respectively. Then, the metallicity observed at each location $\Tilde{Z}(\vec x)$ after being smeared by a point spread function with kernel $K(\vec x)$ is given by:

\begin{equation}
    \Tilde{Z}(\vec x) = \frac{\int Z(\vec x') I(\vec x') K( \vec x' - \vec x) d \vec x'}{\int I(\vec x') K( \vec x' - \vec x) d \vec x'}.
    \label{eq:PSF_smear}
\end{equation}

To find the metallicity measured within an aperture $A$, we must then take a light-weighted average of this quantity over a region:

\begin{equation}
    \langle \Tilde{Z} \rangle_A = \frac{\int_A \Tilde{Z}(\vec x) \Tilde{I}(\vec x) d \vec x}{\int_A \Tilde{I}(\vec x) d \vec x}.
    \label{eq:pixellisation}
\end{equation}

Here, $\Tilde{I}(\vec x)$ is the PSF-smeared surface brightness at each location $\vec x$ within the aperture $A$.

Our new tracer module implemented in \textsc{lenstronomy} accounts for both of these effects, in order to properly integrate the effects of beam smearing and pixelisation of light-weighted quantities into a forward-modelling framework. 

Additionally, we also allow the light-weighted metallicity from different components to be blended together, to model galaxies that are not well described by a single light profile. For a model of a galaxy with multiple surface brightness profiles $I_1(\vec x), \dots, I_n(\vec x)$, each with their own unique metallicity profiles $Z_1(\vec x), \dots, Z_n(\vec x)$, we may compute the intrinsic metallicity profile produced by a sum of these components as: 

\begin{equation}
    Z(\vec x) = \frac{\sum_{i=1}^{n} I_i(\vec x)Z_i(\vec x) }{\sum_{i=1}^{n} I_i(\vec x)} .
    \label{eq:summing}
\end{equation}

This metallicity profile can then be smeared by the point spread function (Equation \ref{eq:PSF_smear}) and pixelised (Equation \ref{eq:pixellisation}) to model the metallicity that we would measure within each pixel. This process allows us to model families of metallicity distributions that are not radially symmetric, which may be more suitable for describing peculiar or merging systems.

\begin{figure*}
    \centering
    \includegraphics[width=0.8\textwidth]{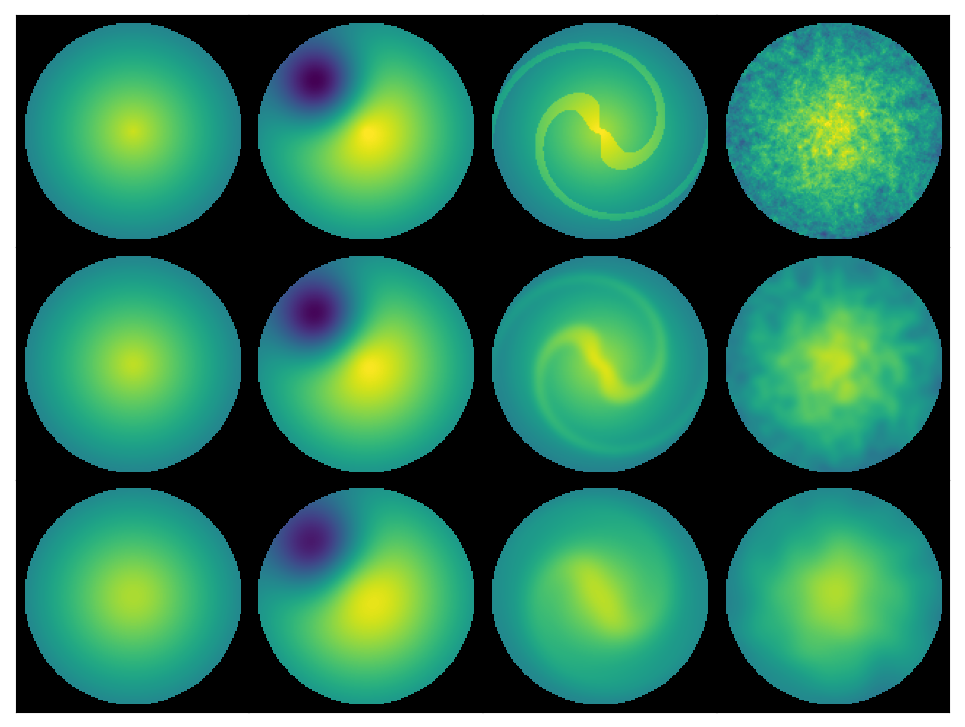}
    \caption{In the top row, we show the four toy metallicity distributions shown in Figure \ref{fig:same_gradZ_diff_galaxies}, with no PSF smear. Below these models, we show the light-weighted properties of these galaxies smeared by a Gaussian PSF with moderate (\textit{middle row}) and poor resolution (\textit{bottom row}). As the resolution becomes increasingly coarse, certain details of galaxy structure are lost. We show this effect quantitatively in Figure \ref{fig:vs_psf}.}
    \label{fig:smearing}
\end{figure*}

We show examples of how these equations affect observed metallicity distributions of galaxies in Figure \ref{fig:smearing}. Modelling the light distribution of all of these galaxies as an exponential, $\Tilde{I}(\vec x) \propto \exp ( - ||\vec x|| / R_e )$, we show the effects of a light-weighted convolution of our four toy model galaxies introduced above with a Gaussian PSF kernel of two different widths. The first of these (middle row of Figure \ref{fig:smearing}) has a PSF FWHM of $\sim 0.6 R_e$, and the second (bottom row) has a PSF FWHM of $\sim 1.3 R_e$ corresponding to cases where the resolution is moderate or poor, respectively. We see that at moderate resolution, fine details on the small-scale structure of galaxies are buried, but other non-axisymmetric spatial trends associated with these galaxies can still be recovered. At poor resolution, spatial features such as spiral arms become washed out, but large-scale variations in the metallicity profile of the galaxy can still be seen, such as the recent merger with a metal-poor companion shown in the second column. This demonstrates how (i) the effects of PSF smearing must be accounted for to properly model a galaxy's metallicity distribution (we make this argument quantitatively in Section \ref{sec:gradient}), and (ii) even in the presence of large PSF smear, there are opportunities to fit multi-component 2-dimensional models to observed metallicity distributions in order to capture astrophysical information about the evolution history of these galaxies.

While this methodology was designed for resolved metallicity data, it can be equally well applied for any light-weighted quantity that can be spatially resolved over a galaxy, such as its velocity distribution, or the colour observed in each spaxel. This module was released as a part of \lenstronomy\ package in Version 1.11.6, available through PyPI and the anaconda distribution. A tutorial on its use for fitting metallicity profiles is also available online.\footnote{\url{github.com/astrobenji/lenstronomy-metals-notebooks}}

\section{Use case \#1: Unbiasing metallicity gradients computed from PSF-smeared data}
\label{sec:gradient}

\citet{Acharyya+20} showed that when the size of a PSF becomes comparable to the scale radius of a galaxy, the metallicity gradients recovered using standard techniques were biased to recover flatter values for the metallicity gradient. In this Section, we demonstrate that the forward-modelling approach presented in this work and implemented in\textsc{lenstronomy} is able to account for the effect of PSF smearing and accurately recover the underlying metallicity gradients of galaxies, even when the effects of PSF smearing grow very large.

To demonstrate this, we simulated observations of a simulated galaxy with an exponential light profile, with an effective radius of $R_e = 1$", typical of a spiral galaxy \citep[e.g.][]{vanderWel+24}. We did not account for any inclination effects, assuming this galaxy is observed face on. The metallicity profile of this galaxy was chosen to have a central metallicity of $12 + \log([$O/H$])=8$ and a strong metallicity gradient of $-0.2$ dex per arcsecond.

We simulated observations of this galaxy using Gaussian PSFs of 8 different widths, with full widths at half maxima (FWHMs) $\in \{0.1,0.2,0.3,0.5,0.7,1.0,1.5,2.0\}$ arcsec. Using our light-weighed forward modelling approach implemented in \textsc{lenstronomy}, we simulated observations of this galaxy with a field of view of $8'' \times 8''$, to capture the outskirts of this galaxy. The size of each pixel was taken to be $0.066''$, to match the pixel size of \textit{JWST}'s NIRISS instrument. In this experiment, the size of a pixel was kept constant as the PSF was changed. 

Uncertainty in the metallicity observed for each pixel is set by the signal to noise of the different emission lines in each pixel. We model this as following $\sigma_Z^2 = c_1 + \frac{c_2}{I}$, where $I$ is the intensity of light in each pixel (see Appendix \ref{ap:heteronoise} for the physical motivation behind this noise model). We fit the constants $c_1$ and $c_2$ such that $\sigma_Z^2=0.01$ at the centre of the galaxy, and $\sigma_Z^2=0.02$ at a distance of $1R_e$ from the galaxy's centre. These values were chosen based on typical uncertainties found in resolved metallicity maps from gravitationally lensed observations \citep[e.g.][]{Wang+20}.

Using this collection of simulated galaxy observations, we then tested our ability to recover the galaxy's metallicity gradient in two different ways. First, we used a standard weighted least-squares (WLS) approach, ignoring the presence of a PSF and weighting the metallicity recovered in each spaxel by its inverse variance, as implemented through the Python package \textsc{scikit-learn}. In this procedure, the central metallicity, $Z_c$, and the metallicity gradient, $\gradZ$, of a galaxy are estimated from a collection of metallicities $\{ Z_i \}$ that are measured at physical distances $\{ r_i \}$ from the galaxy's centre. Error on measured values of $\{ r_i \}$ are treated as negligible, while each metallicity measurement has its own measurement uncertainty $\{ \sigma_i \}$. Then, best-fit values for the two parameters $Z_c$ and $\gradZ$ are computed using chi-squared minimisation:

\begin{equation}
    \chi^2 = \sum_i \frac{(Z_i - Z_c - \gradZ \cdot r_i)^2}{2\sigma_i^2}.
\end{equation}

Secondly, we used our updated version of \textsc{lenstronomy}, incorporating the effects of PSF smearing to recover the metallicity gradients via a forward-modelling approach. Specifically, we used a particle swarm optimisation algorithm \citep{PSO} to find the initial conditions for a MCMC approach using \textsc{emcee} \citep{emcee}, then ran \textsc{emcee} using 100 walkers and 250 steps, discarding 50 steps as a burn-in stage. We compute the median value of $\nabla Z$ and a $68\%$ credible interval by looking at the distribution of samples fit by \textsc{emcee}, following the recommendations of the package's developers \citep{Hogg+Foreman-Mackey18}.

\begin{figure}
    \centering
    \includegraphics[width=0.5\textwidth]{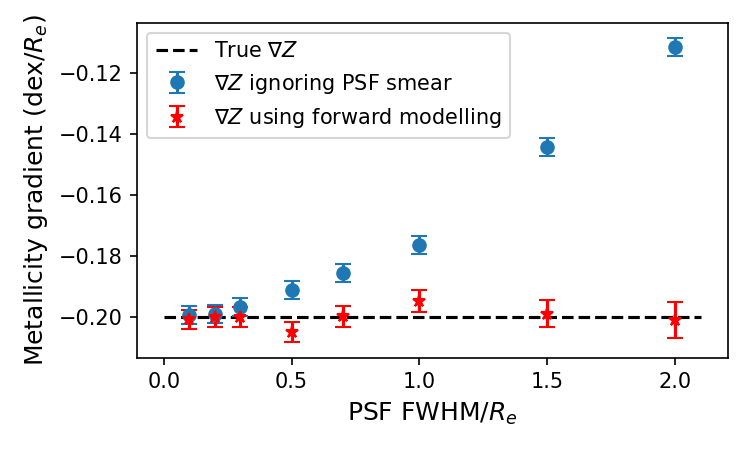}
    \caption{Recovered values of the metallicity gradient as a function of the size of the point spread function, using the classical weighted least-squares approach without forward-modelling (blue), and using \textsc{lenstronomy}'s implementation of a forward-modelling approach (red). When the size of the PSF, measured using its FWHM, becomes comparable to the size of the galaxy, the metallicity gradient recovered by the traditional approach becomes biased to values closer than zero. However, forward modelling can be used to eliminate this biasing effect.}
    \label{fig:vs_psf}
\end{figure}

In Figure \ref{fig:vs_psf}, we plot the best-fit values of $\nabla Z$ computed for each simulated galaxy observation, with each of these two data pipelines. We see that, when a WLS approach is used with no PSF correction, the recovered metallicity gradient becomes flat when the with of the PSF becomes comparable to the radius of the galaxy (FWHM$/R_e \geq 0.7$). This bias is significant, and the effect grows stronger when the beam is made wider. On the other hand, when \textsc{lenstronomy} is used to forward model the effects of PSF smearing, no such bias is seen, even when the FWHM of the PSF becomes much larger than the scale radius of the galaxy. Six out of eight MCMC trials contain the true value of $\nabla Z$ within their $68\%$ credible intervals, and all contain the true value of $\nabla Z$ within a $95\%$ credible interval (not shown), which is the statistically expected result. This indicates that a forward-modelling approach that accounts for the effects of PSF smearing is required to get unbiased results, such as the method implemented in \lenstronomy.

While forward-modelling is a powerful approach for recovering spatial information from PSF-smeared data, its performance is still sensitive to resolution.
Forward modelling cannot be used to recover any information on spatial scales that are smaller than a single pixel. 
Whereas the effects of PSF smearing can be modelled and corrected for, the data will not contain any information on processes that occur on spatial scales smaller than the scale at which the data is sampled \citep{cressie93}. This represents a fundamental limit on the quality of data for which models of the metallicity distribution can be constructed.

To demonstrate this fundamental limitation, we explored how the metallicity gradients recovered using \textsc{lenstronomy} are affected by the size of the galaxy in question. For this exploration, we fix the pixel size to be half the FWHM of the PSF, simulating Nyquist sampling, and explore galaxies with effective radii that extend over $1, 1.4, 2,3,5,$ and $8$ pixels, corresponding to beam widths of $2, 1.4, 1, 0.66, 0.4,$ and $0.25$ FWHM$/R_e$. In each case, we set the field of view of each observation to be up to $4R_e$ from the galaxy's centre, to ensure the entirety of the galaxy lies within our field of view. 

\begin{figure}
    \centering
    \includegraphics[width=0.5\textwidth]{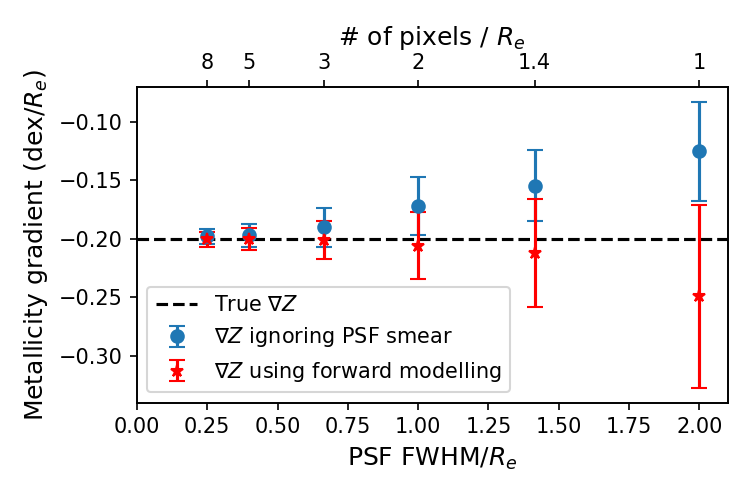}
    \caption{The effect of the galaxy size on the performance of metallicity gradient fitting using two methods: the standard approach that neglects PSF smearing (blue), and our forward-modelling approach (red). For marginally-resolved galaxies  with $\sim 1-2$ pixels per $R_e$, a forward-modelling approach yields more reliable metallicity gradient estimates. For small galaxies with $R_e \lesssim 1$ pixels, there is no longer enough information in the data to precisely fit the metallicity gradient using either method.}
    \label{fig:vs_size}
\end{figure}

To assess whether any bias is seen in the recovered metallicity gradients using \textsc{lenstronomy}, we fit the metallicity profile using each experimental setup 100 times, regenerating the noise from the simulated heteroskedastic metallicity variance maps for each trial. As a control, we repeat this using the WLS method without accounting for PSF smearing.
We show the $16$th, $50$th (median) and $84$th percentiles in the metallicities recovered at each resolution in Figure \ref{fig:vs_size}. We find that both methods become unreliable when the size of a pixel becomes comparable to the $R_e$ of the galaxy, but \lenstronomy\ produces results that are more consistent with the true metallicity gradient at all resolutions. In particular, when the galaxy extends over only a few pixels, the uncertainty on the metallicity gradient recovered by \textsc{lenstronomy} becomes large too, successfully capturing the lack of information in the data.

Such an effect of the uncertainty on the metallicity profile has been noted before, e.g. by \citet{Belfiore+17}. For this reason, longitudinal studies that seek to capture radial metallicity profiles for a wide variety of galaxies in the local Universe are limited to galaxies of a certain size, set by the beam width of the telescopes in use. In \citet{Sanchez+21}, in order to ensure the quality of their data, they restrict their analysis to galaxies with PSF FWHM$/R_e < 0.5$. \citet{Barrera-Ballesteros+23} take a similar approach, but use a different limiting galaxy size to ensure that PSF FWHM$/R_e < 1.0$. Our Figure \ref{fig:vs_size} shows that, when a forward modelling approach is used to correct the metallicity profiles for the effects of PSF smear, a looser criterion of FWHM$/R_e < 1.5$ is required for metallicity gradients to be recovered accurately. This demonstrates that the use of the methods included in our newly updated version of \lenstronomy\ hold the power to extend the analysis of local galaxies to a larger sample size, allowing the radial profiles of more compact systems that are more reminiscent of high-redshift galaxies to be explored.

\section{Use case \#2: Fitting blended metallicity profiles to clumpy galaxies}
\label{sec:clumpy}

Most galaxies at cosmic noon are not disk galaxies \citep[e.g.][]{Conselice+08}. Most galaxies at cosmic noon do not show a well-established metallicity gradient \citep{Simons+21}. To understand the chemical structure of these galaxies, more complex models of the chemical morphology of galaxies must be considered.
 
\begin{figure*}
    \centering
    \includegraphics[width=0.95\textwidth]{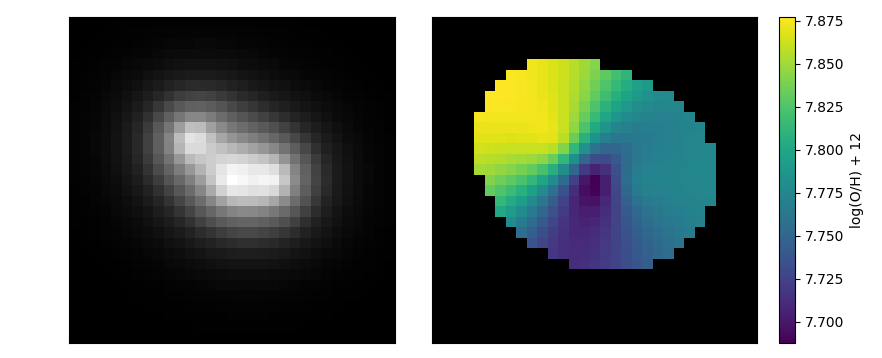}
    \caption{The light profile (left) and metallicity profile (right) for a simulated galaxy with an irregular morphology and a non-linear metallicity profile. This galaxy is made of three different components, all of which are modelled to have exponential light profiles with the same brightness and the same $R_e$. Each of the three components of this galaxy is modelled as having a constant, but different, metallicity, with the lowest metallicity in the centre, and the highest metallicity in the top-left corner.
    }
    \label{fig:clumpy_sim}
\end{figure*}

To demonstrate the ability to use \textsc{lenstronomy} to fit metallicity profiles other than a linear metallicity gradient, we construct a simulated setup of a clumpy galaxy, using three identical exponential light profiles with effective radii of $R_e=0.25''$ (shown in the left panel of Figure \ref{fig:clumpy_sim}). The intrinsic metallicity of the central clump is chosen to be $12+\log($O/H$)=7.6$ throughout. The second clump, offset by $0.25''$ in the positive x-direction from the central clump, has an intrinsic metallicity of $12+\log($O/H$)=7.8$, and the final clump, above and to the left of the central clump separated by $0.35''$ is chosen to have a metallicity of $12+\log($O/H$)=7.9$. Such a metallicity structure was chosen as a plausible setup involving three interacting dwarf galaxies coming together at cosmic noon, which would appear to have a positive metallicity gradient when observed using a radially-symmetric model. 

We simulate an observation of this galaxy using the JWST instrument NIRISS, with a resolution of $0.066''$ per pixel, using a simulated PSF of the F150W filter computed using the Python package \textsc{webbpsf}\footnote{Made by the Space Telescope Science Institute, available at \url{https://www.stsci.edu/jwst/science-planning/proposal-planning-toolbox/psf-simulation-tool}.}. As in the previous Section, we model the uncertainty in the metallicity in a heteroskedastic way, letting $\sigma_Z^2=0.01$ at the centres of each clump, increasing to $\sigma_Z^2=0.02$ at a distance of $1R_e$ from each clump. 

We show the metallicity profile of this galaxy, without noise, in the right-hand panel of Figure \ref{fig:clumpy_sim}. We note that we have not simulated any mixing of metallicities between the three components that are interacting to form this system, as the turbulent processes that govern metal mixing in galaxies and their associated timescales are still not well understood. Instead, all gradients in metallicity between components are simply an artefact of the light blending between components and PSF smear.


We attempt to fit the metallicity profile of this galaxy in three different ways. Firstly, we use the WLS method with no PSF correction to fit a radially symmetric metallicity profile to this galaxy. Secondly, we fit a linear gradient using \lenstronomy, accounting for PSF smearing and pixelisation, but ignoring the structure of the galaxy.
Finally, we use \lenstronomy\ to attempt to estimate the intrinsic metallicity of each of the three components that makes up this galaxy. 

\begin{figure}
    \centering
    \includegraphics[width=0.5\textwidth]{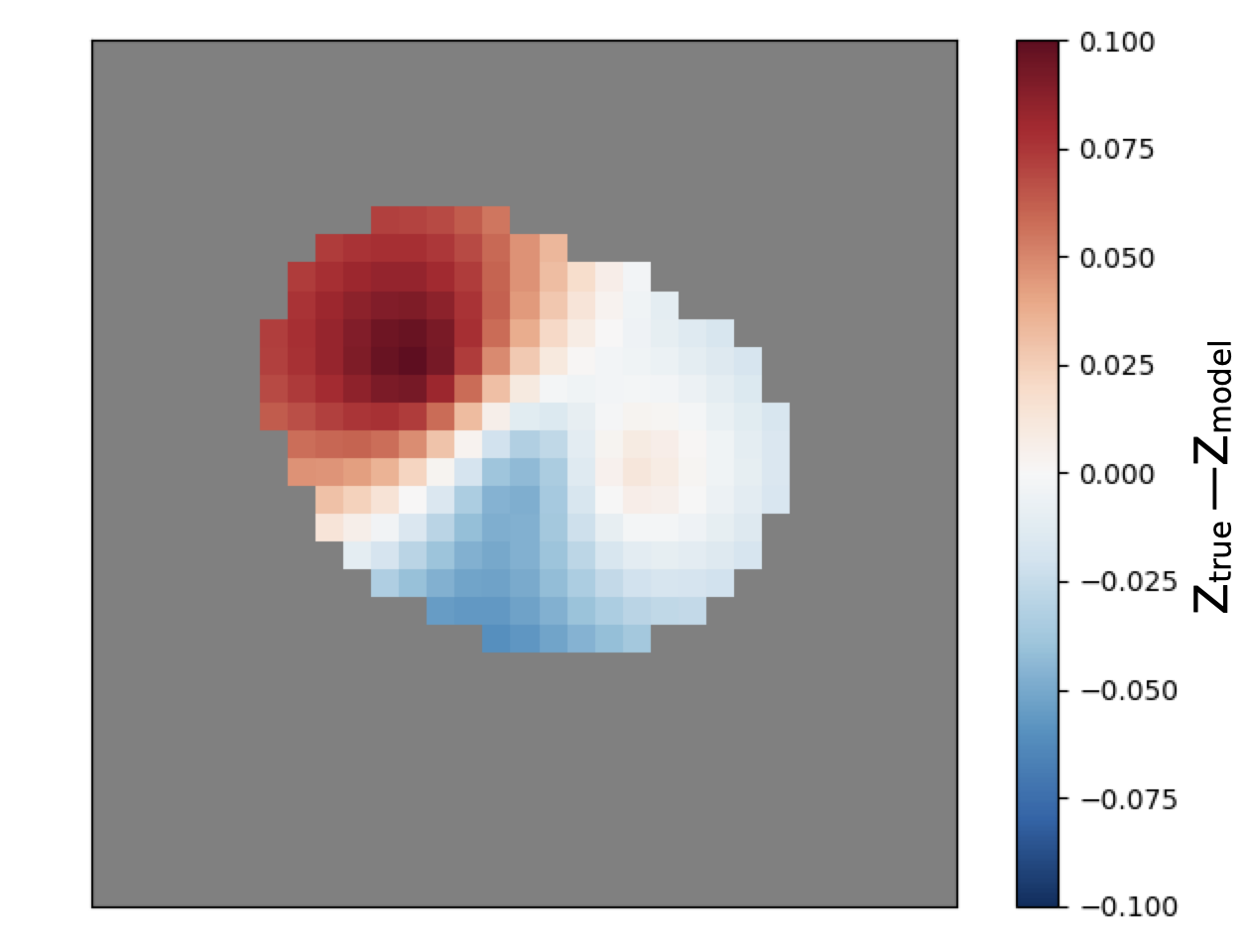}
    \caption{The difference between the metallicity of the galaxy measured within each pixel, and the metallicities predicted using a linear-gradient model. Because this galaxy is not radially symmetric, the metallicity gradient model leaves large residuals of the order 0.1 dex between the observed and recovered metallicities. }
    \label{fig:lingrad_doesnt_fit}
\end{figure}

We find that when a linear gradient is fit, without accounting for pixelisation or PSF smearing, a central metallicity of $12+\log($O/H$)=7.73 \pm 0.02$ with an inverted metallicity gradient of $0.10 \pm 0.03$ dex/arcsec is preferred. This was consistent with the results found by \lenstronomy\ when a linear gradient model was fit, with a central metallicity of $12+\log($O/H$)=7.73 \pm 0.02$ and a metallicity gradient of $0.12 \pm 0.04$ dex/arcsec being favoured. However, neither of these solutions are successful at describing the internal metallicity structure of this galaxy to high accuracy. In Figure \ref{fig:lingrad_doesnt_fit}, we plot the differences between the metallicities predicted by the linear gradient model fit using a least-squares approach and the true simulated metallicity of each spaxel. We find that there exist large, asymmetric residuals between the modelled and observed metallicity profiles with this galaxy, showing deviations of up to $0.1$ dex, even within the brightest regions where the signal-to-noise ratio of the metallicity data was highest.

Finally, we use \lenstronomy\ to attempt to recover the metallicity profile of this galaxy by modelling each clump as having a constant metallicity throughout. We find that, even with noisy metallicity data, if the light profile of the galaxy can be modelled accurately then the metallicity of each of the three clumps could be recovered with a high degree of accuracy, with errors smaller than 0.01 dex (see Table \ref{tab:lenstronomy_blob_fit}). The largest deviation between this model and the data was found to be only $0.0084$ dex, representing a $10$-fold improvement in prediction accuracy over models in which metallicity is assumed to be radially symmetric throughout a galaxy. More formally, using the Bayesian information criterion, the 3-component model is highly preferred to a linear gradient model for this system, with $\Delta$BIC$=35$.

\begin{table}
    \centering
    \begin{tabular}{r|c|c}
     & Metallicity estimate & Input metallicity \\
     Component $\#1$ & $7.6067 \pm  0.032$ & $7.6$ \\
     Component $\#2$ & $7.8089 \pm  0.024$ & $7.8$ \\
     Component $\#3$ & $7.9002^{+0.0192}_{-0.0195}$ & $7.9$ \\
    \end{tabular}
    \caption{Median, 16th, and 84th percentiles of the recovered metallicities of each component of the simulated galaxy. We find that \lenstronomy\ is very successful at accurately recovering the signal, even when the amount of noise is large.}
    \label{tab:lenstronomy_blob_fit}
\end{table}

\section{Discussion}
\label{sec:discussion}

Our approach of using forward-modelling to correct for the effect that PSF-smearing has on flattening metallicity gradients is not unique. A similar approach has been discussed in \citet{Carton+17}, and applied to MUSE data in \citet{Carton+18}. However, this forward-modelling approach differs from our own in several ways. Firstly, their model incorporates a photoionisation model, and so requires as input several emission-line images. On the other hand, the \lenstronomy\ approach takes as input a metallicity map that can be constructed a priori through any method that the observer has access to, making it more flexible. Secondly, the model of \citet{Carton+17} allows resolved metallicity measurements to be binned (i.e. into a Voronoi tessellation), whereas our code does not have this feature. Finally, the model of \citet{Carton+17} assumes that the true metallicity distribution of a galaxy is always a metallicity gradient, while our code allows the user to specify any functional form for the underlying metallicity structure and surface brightness distributions, allowing for asymmetries in a galaxy's chemical profile to be captured. Furthermore, \lenstronomy\ also contains the capability to work on gravitationally-lensed sources, allowing the intrinsic metallicity distributions of such systems to be uncovered in the source plane.

The quality of our data is improving, and the quality of our models ought to improve with it. 
Thanks to \textit{JWST}, we are already collecting IFS data of gravitationally-lensed galaxies at $z\sim 3$ with sub-kpc resolution \citep{Wang+22}. Such exquisite data quality allows models that include local variations in metallicity that depart from radial symmetry to be fit. These non-axisymmetric models allow signatures of gas-driven galaxy evolution, such as minor mergers, recent starbursts, asymmetric outflows, or cold flows of pristine gas into a galaxy's outskirts along cosmic filaments to be recognised in observational data. 

\citet{Tissera+16} examined abundance profiles for galaxies within a cosmological simulation, and found that the galaxies with the steepest positive and negative metallicity gradients showed evidence of substructures such as rings, bars, and close companions, demonstrating a link between abundance profiles and morphology. The techniques that we explore in this work allow us to delve deeper into this relationship in a quantitative way.

Such irregular metallicity profiles have been seen in observational data for high redshift galaxies. \citet{Forster-Schreiber+18} examine the two-dimensional distributions of the line ratio [N \textsc{ii}]/H$\alpha$, which is approximately proportional to metallicity \citep{Pettini+Pagel04}. They find that often, large asymmetries can be seen in the [N \textsc{ii}]/H$\alpha$ profiles, reflecting asymmetries in the surface brightness profiles of these systems. A similar effect is commented on in \citet{Curti+20b}. In their Figure 11, they show four galaxies observed at high redshift with large asymmetries in their metallicity profiles, revealing clumpy substructures which could be a sign of ongoing interactions within a merging system, or different phenomena such as gas flows that act on timescales shorter than secular processes. 

When a radial gradient alone is used to characterise all metallicity variation within high-redshift galaxies, radially-symmetric physical models are employed to explain this variation. Of the three high-redshift galaxies exhibiting inverted metallicity gradients presented in \citet{Cresci+10}, two show visible departures from radial symmetry in their 2D metallicity profiles (their Figure 1). However, as this variation was not captured, all three radial metallicity profiles were explained using the same model of cold gas flows being preferentially directed into a galaxy's centre \citep{Keres+05}. \citet{Schonrich+McMillan17} argue that this model may not be sufficient to explain the inverted gradients captured within this data, as the effects of enhanced star formation efficiency in a galaxy's central regions may increase the metallicity of the central regions more than could be compensated for by a pristine gas flow. Fitting more flexible models to the data has the potential to reveal alternative explanations for such inverted, irregular metallicity profiles.

The systems with large asymmetries in their chemical profiles reported in \citet{Curti+20b} also show large deviations from radial symmetry in their photometric and kinematic data. This new version of \lenstronomy\ allows us to account for information present in these ancillary data product, to inform us of the presence of substructures within the galaxy when fitting more complex metallicity profiles. This gives us a more holistic approach to describe the chemical structure of galaxies, combining data from different instruments and surveys to produce a more complete description of the dominant physical processes that govern galaxy evolution at high redshift.

\subsection{Potential applications to other galaxy properties}

Many different publically-available code bases exist that use forward modelling to accurately fit light profiles to galaxy data, including the effects of pixelisation and PSF smearing, such as \textsc{GALFIT} \citep{galfit} and \textsc{galight} \citep{galight}. 
This work presents an extension of \lenstronomy\ that uses forward-modelling to compute the effects of gravitational lensing, PSF smearing, and pixelisation on the spatial distribution of \emph{light-weighted properties} of a galaxy. This paper investigates the performance of this code base on accurately resolving the metallicity distributions of galaxies. However, the algorithms we have developed will also work on any other light-weighted property of galaxies. For example, forward-modelling of elemental abundance ratios such as [C/O] would allow us to see signatures of recent star formation, as $\alpha$ elements such as Oxygen are primarily released in core-collapse supernovae with delay times of $\sim10$ Myr, whereas a significant amount of Carbon is additionally released during the end of life of intermediate mass stars, with delay times closer to $\sim 100$ Myr$-10$ Gyr \citep{Kobayashi+11, Kobayashi+20, Jones+23}. However, such observations would be complicated by details of turbulent metal-mixing in these high-redshift galaxies, which is not well understood \citep{Metha+21}.

Alternatively, the age distribution of stars throughout a galaxy could be constrained via SED fitting. Such an analysis has already been attempted by \citet{Gimenez-Arteaga+23}. By fitting a SED independently to each pixel of a galaxy observed by NIRCam, several properties, including the age, dust attenuation, and star formation rate density of the galaxy could be recovered for each pixel. However, the effects of PSF smearing through pixels of different light sensitivity were not accounted for in this analysis, limiting their interpretation of resolved variables to scales larger than a PSF beam. Using \lenstronomy, these final age maps could be combined with light profile data in order to recover the intrinsic age distribution of stars within these galaxies at native resolution, perhaps allowing a detailed analysis on the properties of small star-forming clumps to be performed on high redshift galaxies. 

Another natural application of this forward modelling approach would be to analyse kinematic structures in galaxies captured by IFS data. Like metallicity, velocity dispersion is a light-weighted quantity. A similar approach has already been used by \citet{Graaff+23} to uncover kinematic profiles of six galaxies at $5.5 < z < 7.6$ observed with JWST NIRSpec data, using a bespoke forward-modelling pipeline that is specifically designed to account for the effects of the microshutter arrays used for multi-object spectroscopy (MOS). The new module implemented in \lenstronomy\ is less suitable for NIRSpec MOS data than the code that they present, but would be suitable for observation campaigns using more typical IFS instruments, such as NIRISS or MUSE. 

\section{Summary}
\label{sec:summary}

Spatially-resolved observations of light-weighted properties such as metallicity can be heavily impacted by instrumental effects, such as PSF smearing and pixelisation.
In this work, we present an added module in the Python package \lenstronomy, and demonstrate its ability to correct for these effects, as well as fit flexible models of the chemical structure of galaxies, through forward-modelling. We summarise our main conclusions below:

\begin{itemize}
    \item We explored the ability of \lenstronomy\ to recover the metallicity profile of a galaxy under the effects of PSF smearing. We found that, when the size of a pixel was small compared to the size of a galaxy, the new \lenstronomy\ code was able to recover the true metallicity gradients of a simulated galaxy (observed with realistic noise) without bias, even when the size of the PSF became comparable to the size of the galaxy itself. This makes this code very useful for situations where an instrument has a large PSF that extends over many pixels.
    \item We additionally tested the performance of \lenstronomy\ over Nyquist-sampled mock-IFS data. We found that accounting for PSF smear and pixelisation using forward-modelling yielded more accurate metallicity gradients than when these effects were ignored for marginally resolved data ($\sim 1.5-3$ pixels per $R_e$), allowing accurate metallicity profiles to be recovered for smaller systems. Data with a resolution of $\sim 1$ pixel per $R_e$ was found to not contain enough information for a metallicity gradient to be fit precisely even when PSF smear is accounted for, leaving large uncertainties on the recovered metallicity profile parameters.
    \item To test the ability of \lenstronomy\ to fit non-linear metallicity profiles of galaxies, we simulated a galaxy composed of three ``clumps" of identical brightness and different metallicities. We found that, while traditional model-fitting pipelines would characterise this galaxy as having an inverted metallicity gradient, \lenstronomy was able to successfully resolve the metallicities of the three components with high accuracy.
\end{itemize}

These tests demonstrate the utility of \lenstronomy\ to capture accurately details of a galaxy's chemical profile in two-dimensions, even when the seeing is poor. This is particularly relevant at high-redshift, where a multitude of processes are expected to cause large deviations from a circularly-symmetric galaxy profile. In future works, we will apply these techniques to resolved maps of gravitationally-lensed galaxies captured by NIRISS in the GLASS-JWST survey (PI:Treu). As with the rest of the \lenstronomy\ package, all of our code is actively maintained and documented, is available for public use under the MIT license and is distributed through the python packaging index.

\section*{Acknowledgements}
We thank the referee of this paper for their attentive reading and constructive feedback, which improved the quality of this work.
This research is supported in part by the Australian Research Council Centre of Excellence for All Sky Astrophysics in 3 Dimensions (ASTRO 3D), through project number CE170100013.
BM acknowledges support from Australian Government Research Training Program (RTP) Scholarships and the Jean E Laby Foundation.
BM would like to thank the staff and students of the University of California, Los Angeles, where the majority of this research was conducted, for their assistance in facilitating a safe and successful study trip. BM thanks Nithin Babu for being an early tester of the Jupyter notebook tutorial, and Alex Clark for a stimulating lunchtime conversation. X. W. is supported by the Fundamental Research Funds for the Central Universities, the CAS Project for Young Scientists in Basic Research Grant No. YSBR-062, and the Xiaomi Young Scholars Fellowship.
SB acknowledges support by Stony Brook University. We acknowledge the Gabrielino/Tongva peoples as the traditional owners of Tovaangar (the Los Angeles basin and So. Channel Islands) and the Wurundjeri Woi-wurrung and Bunurong/Boon Wurrung  peoples of the Kulin nation as the traditional owners of Naarm (Melbourne) where this research was primarily conducted.

\section*{Data Availability}



The code is now included as a part of \lenstronomy\ Version 1.12.0 and is available for public download: \url{github.com/lenstronomy/lenstronomy}. Notebooks containing constructions of all results presented in this work will be made available via \url{github.com/astrobenji/lenstronomy-metals-notebooks} upon acceptance of this manuscript.



\bibliographystyle{mnras}
\bibliography{example} 

\begin{thebibliography}{}
\makeatletter
\relax
\def\mn@urlcharsother{\let\do\@makeother \do\$\do\&\do\#\do\^\do\_\do\%\do\~}
\def\mn@doi{\begingroup\mn@urlcharsother \@ifnextchar [ {\mn@doi@}
  {\mn@doi@[]}}
\def\mn@doi@[#1]#2{\def\@tempa{#1}\ifx\@tempa\@empty \href
  {http://dx.doi.org/#2} {doi:#2}\else \href {http://dx.doi.org/#2} {#1}\fi
  \endgroup}
\def\mn@eprint#1#2{\mn@eprint@#1:#2::\@nil}
\def\mn@eprint@arXiv#1{\href {http://arxiv.org/abs/#1} {{\tt arXiv:#1}}}
\def\mn@eprint@dblp#1{\href {http://dblp.uni-trier.de/rec/bibtex/#1.xml}
  {dblp:#1}}
\def\mn@eprint@#1:#2:#3:#4\@nil{\def\@tempa {#1}\def\@tempb {#2}\def\@tempc
  {#3}\ifx \@tempc \@empty \let \@tempc \@tempb \let \@tempb \@tempa \fi \ifx
  \@tempb \@empty \def\@tempb {arXiv}\fi \@ifundefined
  {mn@eprint@\@tempb}{\@tempb:\@tempc}{\expandafter \expandafter \csname
  mn@eprint@\@tempb\endcsname \expandafter{\@tempc}}}

\bibitem[\protect\citeauthoryear{{Acharyya}, {Krumholz}, {Federrath}, {Kewley},
  {Goldbaum}  \& {Sharp}}{{Acharyya} et~al.}{2020}]{Acharyya+20}
{Acharyya} A.,  {Krumholz} M.~R.,  {Federrath} C.,  {Kewley} L.~J.,  {Goldbaum}
  N.~J.,   {Sharp} R.,  2020, \mn@doi [\mnras] {10.1093/mnras/staa1100}, \href
  {https://ui.adsabs.harvard.edu/abs/2020MNRAS.495.3819A} {495, 3819}

\bibitem[\protect\citeauthoryear{{Barrera-Ballesteros}
  et~al.,}{{Barrera-Ballesteros} et~al.}{2023}]{Barrera-Ballesteros+23}
{Barrera-Ballesteros} J.~K.,  et~al., 2023, \mn@doi [\rmxaa]
  {10.22201/ia.01851101p.2023.59.02.06}, \href
  {https://ui.adsabs.harvard.edu/abs/2023RMxAA..59..213B} {59, 213}

\bibitem[\protect\citeauthoryear{{Belfiore} et~al.,}{{Belfiore}
  et~al.}{2017}]{Belfiore+17}
{Belfiore} F.,  et~al., 2017, \mn@doi [\mnras] {10.1093/mnras/stx789}, \href
  {https://ui.adsabs.harvard.edu/abs/2017MNRAS.469..151B} {469, 151}

\bibitem[\protect\citeauthoryear{{Belfiore}, {Vincenzo}, {Maiolino}  \&
  {Matteucci}}{{Belfiore} et~al.}{2019}]{Belfiore+19}
{Belfiore} F.,  {Vincenzo} F.,  {Maiolino} R.,   {Matteucci} F.,  2019, \mn@doi
  [\mnras] {10.1093/mnras/stz1165}, \href
  {https://ui.adsabs.harvard.edu/abs/2019MNRAS.487..456B} {487, 456}

\bibitem[\protect\citeauthoryear{{Belley} \& {Roy}}{{Belley} \&
  {Roy}}{1992}]{Belley+Roy92}
{Belley} J.,  {Roy} J.-R.,  1992, \mn@doi [\apjs] {10.1086/191621}, \href
  {https://ui.adsabs.harvard.edu/abs/1992ApJS...78...61B} {78, 61}

\bibitem[\protect\citeauthoryear{{Berg}, {Skillman}, {Garnett}, {Croxall},
  {Marble}, {Smith}, {Gordon}  \& {Kennicutt}}{{Berg} et~al.}{2013}]{Berg+13}
{Berg} D.~A.,  {Skillman} E.~D.,  {Garnett} D.~R.,  {Croxall} K.~V.,  {Marble}
  A.~R.,  {Smith} J.~D.,  {Gordon} K.,   {Kennicutt} Robert~C. J.,  2013,
  \mn@doi [\apj] {10.1088/0004-637X/775/2/128}, \href
  {https://ui.adsabs.harvard.edu/abs/2013ApJ...775..128B} {775, 128}

\bibitem[\protect\citeauthoryear{{Berg}, {Pogge}, {Skillman}, {Croxall},
  {Moustakas}, {Rogers}  \& {Sun}}{{Berg} et~al.}{2020}]{Berg+20}
{Berg} D.~A.,  {Pogge} R.~W.,  {Skillman} E.~D.,  {Croxall} K.~V.,  {Moustakas}
  J.,  {Rogers} N. S.~J.,   {Sun} J.,  2020, \mn@doi [\apj]
  {10.3847/1538-4357/ab7eab}, \href
  {https://ui.adsabs.harvard.edu/abs/2020ApJ...893...96B} {893, 96}

\bibitem[\protect\citeauthoryear{{Bird}, {Kazantzidis}  \& {Weinberg}}{{Bird}
  et~al.}{2012}]{Bird+12}
{Bird} J.~C.,  {Kazantzidis} S.,   {Weinberg} D.~H.,  2012, \mn@doi [\mnras]
  {10.1111/j.1365-2966.2011.19728.x}, \href
  {https://ui.adsabs.harvard.edu/abs/2012MNRAS.420..913B} {420, 913}

\bibitem[\protect\citeauthoryear{{Birrer} \& {Amara}}{{Birrer} \&
  {Amara}}{2018}]{Birrer+Amara18}
{Birrer} S.,  {Amara} A.,  2018, \mn@doi [Physics of the Dark Universe]
  {10.1016/j.dark.2018.11.002}, \href
  {https://ui.adsabs.harvard.edu/abs/2018PDU....22..189B} {22, 189}

\bibitem[\protect\citeauthoryear{{Birrer} et~al.,}{{Birrer}
  et~al.}{2021}]{Birrer+21}
{Birrer} S.,  et~al., 2021, \mn@doi [The Journal of Open Source Software]
  {10.21105/joss.03283}, \href
  {https://ui.adsabs.harvard.edu/abs/2021JOSS....6.3283B} {6, 3283}

\bibitem[\protect\citeauthoryear{{Boissier} \& {Prantzos}}{{Boissier} \&
  {Prantzos}}{1999}]{Boissier+Prantzos99}
{Boissier} S.,  {Prantzos} N.,  1999, \mn@doi [\mnras]
  {10.1046/j.1365-8711.1999.02699.x}, \href
  {https://ui.adsabs.harvard.edu/abs/1999MNRAS.307..857B} {307, 857}

\bibitem[\protect\citeauthoryear{{Bresolin}}{{Bresolin}}{2011}]{Bresolin+11}
{Bresolin} F.,  2011, \mn@doi [\apj] {10.1088/0004-637X/730/2/129}, \href
  {https://ui.adsabs.harvard.edu/abs/2011ApJ...730..129B} {730, 129}

\bibitem[\protect\citeauthoryear{{Bresolin} \& {Kennicutt}}{{Bresolin} \&
  {Kennicutt}}{2015}]{Bresolin+Kennicutt15}
{Bresolin} F.,  {Kennicutt} R.~C.,  2015, \mn@doi [\mnras]
  {10.1093/mnras/stv2245}, \href
  {https://ui.adsabs.harvard.edu/abs/2015MNRAS.454.3664B} {454, 3664}

\bibitem[\protect\citeauthoryear{{Bresolin}, {Ryan-Weber}, {Kennicutt}  \&
  {Goddard}}{{Bresolin} et~al.}{2009}]{Bresolin+09}
{Bresolin} F.,  {Ryan-Weber} E.,  {Kennicutt} R.~C.,   {Goddard} Q.,  2009,
  \mn@doi [\apj] {10.1088/0004-637X/695/1/580}, \href
  {https://ui.adsabs.harvard.edu/abs/2009ApJ...695..580B} {695, 580}

\bibitem[\protect\citeauthoryear{{Bundy} et~al.,}{{Bundy} et~al.}{2015}]{MANGA}
{Bundy} K.,  et~al., 2015, \mn@doi [\apj] {10.1088/0004-637X/798/1/7}, \href
  {https://ui.adsabs.harvard.edu/abs/2015ApJ...798....7B} {798, 7}

\bibitem[\protect\citeauthoryear{{Cameron} et~al.,}{{Cameron}
  et~al.}{2021}]{Cameron+21}
{Cameron} A.~J.,  et~al., 2021, \mn@doi [\apjl] {10.3847/2041-8213/ac18ca},
  \href {https://ui.adsabs.harvard.edu/abs/2021ApJ...918L..16C} {918, L16}

\bibitem[\protect\citeauthoryear{{Carton} et~al.,}{{Carton}
  et~al.}{2017}]{Carton+17}
{Carton} D.,  et~al., 2017, \mn@doi [\mnras] {10.1093/mnras/stx545}, \href
  {https://ui.adsabs.harvard.edu/abs/2017MNRAS.468.2140C} {468, 2140}

\bibitem[\protect\citeauthoryear{{Carton} et~al.,}{{Carton}
  et~al.}{2018}]{Carton+18}
{Carton} D.,  et~al., 2018, \mn@doi [\mnras] {10.1093/mnras/sty1343}, \href
  {https://ui.adsabs.harvard.edu/abs/2018MNRAS.478.4293C} {478, 4293}

\bibitem[\protect\citeauthoryear{{Cedr{\'e}s}, {Cepa}, {Bongiovanni},
  {Casta{\~n}eda}, {S{\'a}nchez-Portal}  \& {Tomita}}{{Cedr{\'e}s}
  et~al.}{2012}]{Cedres+12}
{Cedr{\'e}s} B.,  {Cepa} J.,  {Bongiovanni} {\'A}.,  {Casta{\~n}eda} H.,
  {S{\'a}nchez-Portal} M.,   {Tomita} A.,  2012, \mn@doi [\aap]
  {10.1051/0004-6361/201219571}, \href
  {https://ui.adsabs.harvard.edu/abs/2012A&A...545A..43C} {545, A43}

\bibitem[\protect\citeauthoryear{{Conselice}}{{Conselice}}{2014}]{Conselice14}
{Conselice} C.~J.,  2014, \mn@doi [\araa]
  {10.1146/annurev-astro-081913-040037}, \href
  {https://ui.adsabs.harvard.edu/abs/2014ARA&A..52..291C} {52, 291}

\bibitem[\protect\citeauthoryear{{Conselice} \& {Arnold}}{{Conselice} \&
  {Arnold}}{2009}]{Conselice+Arnold09}
{Conselice} C.~J.,  {Arnold} J.,  2009, \mn@doi [\mnras]
  {10.1111/j.1365-2966.2009.14959.x}, \href
  {https://ui.adsabs.harvard.edu/abs/2009MNRAS.397..208C} {397, 208}

\bibitem[\protect\citeauthoryear{{Conselice}, {Rajgor}  \& {Myers}}{{Conselice}
  et~al.}{2008}]{Conselice+08}
{Conselice} C.~J.,  {Rajgor} S.,   {Myers} R.,  2008, \mn@doi [\mnras]
  {10.1111/j.1365-2966.2008.13069.x}, \href
  {https://ui.adsabs.harvard.edu/abs/2008MNRAS.386..909C} {386, 909}

\bibitem[\protect\citeauthoryear{{Cresci}, {Mannucci}, {Maiolino}, {Marconi},
  {Gnerucci}  \& {Magrini}}{{Cresci} et~al.}{2010}]{Cresci+10}
{Cresci} G.,  {Mannucci} F.,  {Maiolino} R.,  {Marconi} A.,  {Gnerucci} A.,
  {Magrini} L.,  2010, \mn@doi [\nat] {10.1038/nature09451}, \href
  {https://ui.adsabs.harvard.edu/abs/2010Natur.467..811C} {467, 811}

\bibitem[\protect\citeauthoryear{Cressie}{Cressie}{1993}]{cressie93}
Cressie N.,  1993, Statistics for Spatial Data, revised edn.
Wiley, New York

\bibitem[\protect\citeauthoryear{{Croom} et~al.,}{{Croom} et~al.}{2012}]{SAMI}
{Croom} S.~M.,  et~al., 2012, \mn@doi [\mnras]
  {10.1111/j.1365-2966.2011.20365.x}, \href
  {https://ui.adsabs.harvard.edu/abs/2012MNRAS.421..872C} {421, 872}

\bibitem[\protect\citeauthoryear{{Croxall}, {Pogge}, {Berg}, {Skillman}  \&
  {Moustakas}}{{Croxall} et~al.}{2015}]{Croxall+15}
{Croxall} K.~V.,  {Pogge} R.~W.,  {Berg} D.~A.,  {Skillman} E.~D.,
  {Moustakas} J.,  2015, \mn@doi [\apj] {10.1088/0004-637X/808/1/42}, \href
  {https://ui.adsabs.harvard.edu/abs/2015ApJ...808...42C} {808, 42}

\bibitem[\protect\citeauthoryear{{Croxall}, {Pogge}, {Berg}, {Skillman}  \&
  {Moustakas}}{{Croxall} et~al.}{2016}]{Croxall+16}
{Croxall} K.~V.,  {Pogge} R.~W.,  {Berg} D.~A.,  {Skillman} E.~D.,
  {Moustakas} J.,  2016, \mn@doi [\apj] {10.3847/0004-637X/830/1/4}, \href
  {https://ui.adsabs.harvard.edu/abs/2016ApJ...830....4C} {830, 4}

\bibitem[\protect\citeauthoryear{{Curti}, {Mannucci}, {Cresci}  \&
  {Maiolino}}{{Curti} et~al.}{2020a}]{Curti+20}
{Curti} M.,  {Mannucci} F.,  {Cresci} G.,   {Maiolino} R.,  2020a, \mn@doi
  [\mnras] {10.1093/mnras/stz2910}, \href
  {https://ui.adsabs.harvard.edu/abs/2020MNRAS.491..944C} {491, 944}

\bibitem[\protect\citeauthoryear{{Curti} et~al.,}{{Curti}
  et~al.}{2020b}]{Curti+20b}
{Curti} M.,  et~al., 2020b, \mn@doi [\mnras] {10.1093/mnras/stz3379}, \href
  {https://ui.adsabs.harvard.edu/abs/2020MNRAS.492..821C} {492, 821}

\bibitem[\protect\citeauthoryear{{Dekel} et~al.,}{{Dekel}
  et~al.}{2009}]{Dekel+09}
{Dekel} A.,  et~al., 2009, \mn@doi [\nat] {10.1038/nature07648}, \href
  {https://ui.adsabs.harvard.edu/abs/2009Natur.457..451D} {457, 451}

\bibitem[\protect\citeauthoryear{{Ding}, {Silverman}, {Birrer}, {Treu}, {Tang},
  {Yang}  \& {Bottrell}}{{Ding} et~al.}{2022}]{galight}
{Ding} X.,  {Silverman} J.,  {Birrer} S.,  {Treu} T.,  {Tang} S.,  {Yang} L.,
  {Bottrell} C.,  2022, {GaLight: 2D modeling of galaxy images}, Astrophysics
  Source Code Library, record ascl:2209.011 (\mn@eprint {ascl} {2209.011})

\bibitem[\protect\citeauthoryear{{Fathi}, {Beckman}, {Zurita}, {Rela{\~n}o},
  {Knapen}, {Daigle}, {Hernandez}  \& {Carignan}}{{Fathi}
  et~al.}{2007}]{Fathi+07}
{Fathi} K.,  {Beckman} J.~E.,  {Zurita} A.,  {Rela{\~n}o} M.,  {Knapen} J.~H.,
  {Daigle} O.,  {Hernandez} O.,   {Carignan} C.,  2007, \mn@doi [\aap]
  {10.1051/0004-6361:20066990}, \href
  {https://ui.adsabs.harvard.edu/abs/2007A&A...466..905F} {466, 905}

\bibitem[\protect\citeauthoryear{{Foreman-Mackey}, {Hogg}, {Lang}  \&
  {Goodman}}{{Foreman-Mackey} et~al.}{2013}]{emcee}
{Foreman-Mackey} D.,  {Hogg} D.~W.,  {Lang} D.,   {Goodman} J.,  2013, \mn@doi
  [\pasp] {10.1086/670067}, \href
  {https://ui.adsabs.harvard.edu/abs/2013PASP..125..306F} {125, 306}

\bibitem[\protect\citeauthoryear{{F{\"o}rster Schreiber} et~al.,}{{F{\"o}rster
  Schreiber} et~al.}{2018}]{Forster-Schreiber+18}
{F{\"o}rster Schreiber} N.~M.,  et~al., 2018, \mn@doi [\apjs]
  {10.3847/1538-4365/aadd49}, \href
  {https://ui.adsabs.harvard.edu/abs/2018ApJS..238...21F} {238, 21}

\bibitem[\protect\citeauthoryear{{Franchetto} et~al.,}{{Franchetto}
  et~al.}{2021}]{Franchetto+21}
{Franchetto} A.,  et~al., 2021, \mn@doi [\apj] {10.3847/1538-4357/ac2510},
  \href {https://ui.adsabs.harvard.edu/abs/2021ApJ...923...28F} {923, 28}

\bibitem[\protect\citeauthoryear{{Gibson}, {Pilkington}, {Brook}, {Stinson}  \&
  {Bailin}}{{Gibson} et~al.}{2013}]{Gibson+13}
{Gibson} B.~K.,  {Pilkington} K.,  {Brook} C.~B.,  {Stinson} G.~S.,   {Bailin}
  J.,  2013, \mn@doi [\aap] {10.1051/0004-6361/201321239}, \href
  {https://ui.adsabs.harvard.edu/abs/2013A&A...554A..47G} {554, A47}

\bibitem[\protect\citeauthoryear{{Gim{\'e}nez-Arteaga}
  et~al.,}{{Gim{\'e}nez-Arteaga} et~al.}{2023}]{Gimenez-Arteaga+23}
{Gim{\'e}nez-Arteaga} C.,  et~al., 2023, \mn@doi [\apj]
  {10.3847/1538-4357/acc5ea}, \href
  {https://ui.adsabs.harvard.edu/abs/2023ApJ...948..126G} {948, 126}

\bibitem[\protect\citeauthoryear{{Goddard}, {Bresolin}, {Kennicutt},
  {Ryan-Weber}  \& {Rosales-Ortega}}{{Goddard} et~al.}{2011}]{Goddard+11}
{Goddard} Q.~E.,  {Bresolin} F.,  {Kennicutt} R.~C.,  {Ryan-Weber} E.~V.,
  {Rosales-Ortega} F.~F.,  2011, \mn@doi [\mnras]
  {10.1111/j.1365-2966.2010.17990.x}, \href
  {https://ui.adsabs.harvard.edu/abs/2011MNRAS.412.1246G} {412, 1246}

\bibitem[\protect\citeauthoryear{{Grand}, {Kawata}  \& {Cropper}}{{Grand}
  et~al.}{2015}]{Grand+15}
{Grand} R. J.~J.,  {Kawata} D.,   {Cropper} M.,  2015, \mn@doi [\mnras]
  {10.1093/mnras/stv016}, \href
  {https://ui.adsabs.harvard.edu/abs/2015MNRAS.447.4018G} {447, 4018}

\bibitem[\protect\citeauthoryear{{Grand} et~al.,}{{Grand}
  et~al.}{2016}]{Grand+16}
{Grand} R. J.~J.,  et~al., 2016, \mn@doi [\mnras] {10.1093/mnrasl/slw086},
  \href {https://ui.adsabs.harvard.edu/abs/2016MNRAS.460L..94G} {460, L94}

\bibitem[\protect\citeauthoryear{{Grasha} et~al.,}{{Grasha}
  et~al.}{2022}]{Grasha+22}
{Grasha} K.,  et~al., 2022, \mn@doi [\apj] {10.3847/1538-4357/ac5ab2}, \href
  {https://ui.adsabs.harvard.edu/abs/2022ApJ...929..118G} {929, 118}

\bibitem[\protect\citeauthoryear{{Hemler} et~al.,}{{Hemler}
  et~al.}{2021}]{Hemler+21}
{Hemler} Z.~S.,  et~al., 2021, \mn@doi [\mnras] {10.1093/mnras/stab1803}, \href
  {https://ui.adsabs.harvard.edu/abs/2021MNRAS.506.3024H} {506, 3024}

\bibitem[\protect\citeauthoryear{{Ho}, {Kudritzki}, {Kewley}, {Zahid},
  {Dopita}, {Bresolin}  \& {Rupke}}{{Ho} et~al.}{2015}]{Ho+15}
{Ho} I.~T.,  {Kudritzki} R.-P.,  {Kewley} L.~J.,  {Zahid} H.~J.,  {Dopita}
  M.~A.,  {Bresolin} F.,   {Rupke} D. S.~N.,  2015, \mn@doi [\mnras]
  {10.1093/mnras/stv067}, \href
  {https://ui.adsabs.harvard.edu/abs/2015MNRAS.448.2030H} {448, 2030}

\bibitem[\protect\citeauthoryear{{Ho} et~al.,}{{Ho} et~al.}{2017}]{Ho+17}
{Ho} I.~T.,  et~al., 2017, \mn@doi [\apj] {10.3847/1538-4357/aa8460}, \href
  {https://ui.adsabs.harvard.edu/abs/2017ApJ...846...39H} {846, 39}

\bibitem[\protect\citeauthoryear{{Ho} et~al.,}{{Ho} et~al.}{2018}]{Ho+18}
{Ho} I.~T.,  et~al., 2018, \mn@doi [\aap] {10.1051/0004-6361/201833262}, \href
  {https://ui.adsabs.harvard.edu/abs/2018A&A...618A..64H} {618, A64}

\bibitem[\protect\citeauthoryear{{Hogg} \& {Foreman-Mackey}}{{Hogg} \&
  {Foreman-Mackey}}{2018}]{Hogg+Foreman-Mackey18}
{Hogg} D.~W.,  {Foreman-Mackey} D.,  2018, \mn@doi [\apjs]
  {10.3847/1538-4365/aab76e}, \href
  {https://ui.adsabs.harvard.edu/abs/2018ApJS..236...11H} {236, 11}

\bibitem[\protect\citeauthoryear{{Jones}, {Ellis}, {Richard}  \&
  {Jullo}}{{Jones} et~al.}{2013}]{Jones+13}
{Jones} T.,  {Ellis} R.~S.,  {Richard} J.,   {Jullo} E.,  2013, \mn@doi [\apj]
  {10.1088/0004-637X/765/1/48}, \href
  {https://ui.adsabs.harvard.edu/abs/2013ApJ...765...48J} {765, 48}

\bibitem[\protect\citeauthoryear{{Jones} et~al.,}{{Jones}
  et~al.}{2023}]{Jones+23}
{Jones} T.,  et~al., 2023, \mn@doi [\apjl] {10.3847/2041-8213/acd938}, \href
  {https://ui.adsabs.harvard.edu/abs/2023ApJ...951L..17J} {951, L17}

\bibitem[\protect\citeauthoryear{Kennedy \& Eberhart}{Kennedy \&
  Eberhart}{1995}]{PSO}
Kennedy J.,  Eberhart R.,  1995, in Proceedings of ICNN'95 - International
  Conference on Neural Networks. pp 1942--1948 vol.4,
  \mn@doi{10.1109/ICNN.1995.488968}

\bibitem[\protect\citeauthoryear{{Kere{\v{s}}}, {Katz}, {Weinberg}  \&
  {Dav{\'e}}}{{Kere{\v{s}}} et~al.}{2005}]{Keres+05}
{Kere{\v{s}}} D.,  {Katz} N.,  {Weinberg} D.~H.,   {Dav{\'e}} R.,  2005,
  \mn@doi [\mnras] {10.1111/j.1365-2966.2005.09451.x}, \href
  {https://ui.adsabs.harvard.edu/abs/2005MNRAS.363....2K} {363, 2}

\bibitem[\protect\citeauthoryear{{Kewley} \& {Ellison}}{{Kewley} \&
  {Ellison}}{2008}]{Kewley+Ellison08}
{Kewley} L.~J.,  {Ellison} S.~L.,  2008, \mn@doi [\apj] {10.1086/587500}, \href
  {https://ui.adsabs.harvard.edu/abs/2008ApJ...681.1183K} {681, 1183}

\bibitem[\protect\citeauthoryear{{Kewley}, {Geller}  \& {Barton}}{{Kewley}
  et~al.}{2006}]{Kewley+06}
{Kewley} L.~J.,  {Geller} M.~J.,   {Barton} E.~J.,  2006, \mn@doi [\aj]
  {10.1086/500295}, \href
  {https://ui.adsabs.harvard.edu/abs/2006AJ....131.2004K} {131, 2004}

\bibitem[\protect\citeauthoryear{{Kewley}, {Rupke}, {Zahid}, {Geller}  \&
  {Barton}}{{Kewley} et~al.}{2010}]{Kewley+10}
{Kewley} L.~J.,  {Rupke} D.,  {Zahid} H.~J.,  {Geller} M.~J.,   {Barton} E.~J.,
   2010, \mn@doi [\apjl] {10.1088/2041-8205/721/1/L48}, \href
  {https://ui.adsabs.harvard.edu/abs/2010ApJ...721L..48K} {721, L48}

\bibitem[\protect\citeauthoryear{{Kobayashi} \& {Nakasato}}{{Kobayashi} \&
  {Nakasato}}{2011}]{Kobayashi+11}
{Kobayashi} C.,  {Nakasato} N.,  2011, \mn@doi [\apj]
  {10.1088/0004-637X/729/1/16}, \href
  {https://ui.adsabs.harvard.edu/abs/2011ApJ...729...16K} {729, 16}

\bibitem[\protect\citeauthoryear{{Kobayashi}, {Karakas}  \&
  {Lugaro}}{{Kobayashi} et~al.}{2020}]{Kobayashi+20}
{Kobayashi} C.,  {Karakas} A.~I.,   {Lugaro} M.,  2020, \mn@doi [\apj]
  {10.3847/1538-4357/abae65}, \href
  {https://ui.adsabs.harvard.edu/abs/2020ApJ...900..179K} {900, 179}

\bibitem[\protect\citeauthoryear{{Lacey} \& {Fall}}{{Lacey} \&
  {Fall}}{1985}]{Lacey+Fall85}
{Lacey} C.~G.,  {Fall} S.~M.,  1985, \mn@doi [\apj] {10.1086/162970}, \href
  {https://ui.adsabs.harvard.edu/abs/1985ApJ...290..154L} {290, 154}

\bibitem[\protect\citeauthoryear{{Leethochawalit}, {Jones}, {Ellis}, {Stark},
  {Richard}, {Zitrin}  \& {Auger}}{{Leethochawalit}
  et~al.}{2016}]{Leethochawalit+16}
{Leethochawalit} N.,  {Jones} T.~A.,  {Ellis} R.~S.,  {Stark} D.~P.,  {Richard}
  J.,  {Zitrin} A.,   {Auger} M.,  2016, \mn@doi [\apj]
  {10.3847/0004-637X/820/2/84}, \href
  {https://ui.adsabs.harvard.edu/abs/2016ApJ...820...84L} {820, 84}

\bibitem[\protect\citeauthoryear{{L{\'e}pine} et~al.,}{{L{\'e}pine}
  et~al.}{2011}]{Lepine+11}
{L{\'e}pine} J.~R.~D.,  et~al., 2011, \mn@doi [\mnras]
  {10.1111/j.1365-2966.2011.19314.x}, \href
  {https://ui.adsabs.harvard.edu/abs/2011MNRAS.417..698L} {417, 698}

\bibitem[\protect\citeauthoryear{{Li}, {Bresolin}  \& {Kennicutt}}{{Li}
  et~al.}{2013}]{Li+13}
{Li} Y.,  {Bresolin} F.,   {Kennicutt} Robert~C. J.,  2013, \mn@doi [\apj]
  {10.1088/0004-637X/766/1/17}, \href
  {https://ui.adsabs.harvard.edu/abs/2013ApJ...766...17L} {766, 17}

\bibitem[\protect\citeauthoryear{{Li}, {Krumholz}, {Wisnioski}, {Mendel},
  {Kewley}, {S{\'a}nchez}  \& {Galbany}}{{Li} et~al.}{2021}]{Li+21}
{Li} Z.,  {Krumholz} M.~R.,  {Wisnioski} E.,  {Mendel} J.~T.,  {Kewley} L.~J.,
  {S{\'a}nchez} S.~F.,   {Galbany} L.,  2021, \mn@doi [\mnras]
  {10.1093/mnras/stab1263}, \href
  {https://ui.adsabs.harvard.edu/abs/2021MNRAS.504.5496L} {504, 5496}

\bibitem[\protect\citeauthoryear{{Li} et~al.,}{{Li} et~al.}{2023}]{Li+23}
{Li} Z.,  et~al., 2023, \mn@doi [\mnras] {10.1093/mnras/stac3028}, \href
  {https://ui.adsabs.harvard.edu/abs/2023MNRAS.518..286L} {518, 286}

\bibitem[\protect\citeauthoryear{{Ma}, {Hopkins}, {Feldmann}, {Torrey},
  {Faucher-Gigu{\`e}re}  \& {Kere{\v{s}}}}{{Ma} et~al.}{2017}]{Ma+17}
{Ma} X.,  {Hopkins} P.~F.,  {Feldmann} R.,  {Torrey} P.,  {Faucher-Gigu{\`e}re}
  C.-A.,   {Kere{\v{s}}} D.,  2017, \mn@doi [\mnras] {10.1093/mnras/stx034},
  \href {https://ui.adsabs.harvard.edu/abs/2017MNRAS.466.4780M} {466, 4780}

\bibitem[\protect\citeauthoryear{{Maiolino} \& {Mannucci}}{{Maiolino} \&
  {Mannucci}}{2019}]{Maiolino+Mannucci19}
{Maiolino} R.,  {Mannucci} F.,  2019, \mn@doi [\aapr]
  {10.1007/s00159-018-0112-2}, \href
  {https://ui.adsabs.harvard.edu/abs/2019A&ARv..27....3M} {27, 3}

\bibitem[\protect\citeauthoryear{{Marino} et~al.,}{{Marino}
  et~al.}{2012}]{Marino+12}
{Marino} R.~A.,  et~al., 2012, \mn@doi [\apj] {10.1088/0004-637X/754/1/61},
  \href {https://ui.adsabs.harvard.edu/abs/2012ApJ...754...61M} {754, 61}

\bibitem[\protect\citeauthoryear{{Marino} et~al.,}{{Marino}
  et~al.}{2016}]{Marino+16}
{Marino} R.~A.,  et~al., 2016, \mn@doi [\aap] {10.1051/0004-6361/201526986},
  \href {https://ui.adsabs.harvard.edu/abs/2016A&A...585A..47M} {585, A47}

\bibitem[\protect\citeauthoryear{{Martin} \& {Roy}}{{Martin} \&
  {Roy}}{1995}]{Martin+Roy95}
{Martin} P.,  {Roy} J.-R.,  1995, \mn@doi [\apj] {10.1086/175682}, \href
  {https://ui.adsabs.harvard.edu/abs/1995ApJ...445..161M} {445, 161}

\bibitem[\protect\citeauthoryear{{Martin}, {Kobulnicky}  \& {Heckman}}{{Martin}
  et~al.}{2002}]{Martin+02}
{Martin} C.~L.,  {Kobulnicky} H.~A.,   {Heckman} T.~M.,  2002, \mn@doi [\apj]
  {10.1086/341092}, \href
  {https://ui.adsabs.harvard.edu/abs/2002ApJ...574..663M} {574, 663}

\bibitem[\protect\citeauthoryear{{Mast} et~al.,}{{Mast} et~al.}{2014}]{Mast+14}
{Mast} D.,  et~al., 2014, \mn@doi [\aap] {10.1051/0004-6361/201321789}, \href
  {https://ui.adsabs.harvard.edu/abs/2014A&A...561A.129M} {561, A129}

\bibitem[\protect\citeauthoryear{{Metha}, {Trenti}  \& {Chu}}{{Metha}
  et~al.}{2021}]{Metha+21}
{Metha} B.,  {Trenti} M.,   {Chu} T.,  2021, \mn@doi [\mnras]
  {10.1093/mnras/stab2554}, \href
  {https://ui.adsabs.harvard.edu/abs/2021MNRAS.508..489M} {508, 489}

\bibitem[\protect\citeauthoryear{{Metha}, {Trenti}, {Chu}  \&
  {Battisti}}{{Metha} et~al.}{2022}]{Metha+22}
{Metha} B.,  {Trenti} M.,  {Chu} T.,   {Battisti} A.,  2022, \mn@doi [\mnras]
  {10.1093/mnras/stac1484}, \href
  {https://ui.adsabs.harvard.edu/abs/2022MNRAS.514.4465M} {514, 4465}

\bibitem[\protect\citeauthoryear{{Minchev}, {Famaey}, {Combes}, {Di Matteo},
  {Mouhcine}  \& {Wozniak}}{{Minchev} et~al.}{2011}]{Minchev+11}
{Minchev} I.,  {Famaey} B.,  {Combes} F.,  {Di Matteo} P.,  {Mouhcine} M.,
  {Wozniak} H.,  2011, \mn@doi [\aap] {10.1051/0004-6361/201015139}, \href
  {https://ui.adsabs.harvard.edu/abs/2011A&A...527A.147M} {527, A147}

\bibitem[\protect\citeauthoryear{{Moll{\'a}} et~al.,}{{Moll{\'a}}
  et~al.}{2019}]{Molla+19}
{Moll{\'a}} M.,  et~al., 2019, \mn@doi [\mnras] {10.1093/mnras/stz2537}, \href
  {https://ui.adsabs.harvard.edu/abs/2019MNRAS.490..665M} {490, 665}

\bibitem[\protect\citeauthoryear{{Peng}, {Ho}, {Impey}  \& {Rix}}{{Peng}
  et~al.}{2002}]{galfit}
{Peng} C.~Y.,  {Ho} L.~C.,  {Impey} C.~D.,   {Rix} H.-W.,  2002, \mn@doi [\aj]
  {10.1086/340952}, \href
  {https://ui.adsabs.harvard.edu/abs/2002AJ....124..266P} {124, 266}

\bibitem[\protect\citeauthoryear{{P{\'e}rez-Montero}
  et~al.,}{{P{\'e}rez-Montero} et~al.}{2016}]{Perez-Montero+16}
{P{\'e}rez-Montero} E.,  et~al., 2016, \mn@doi [\aap]
  {10.1051/0004-6361/201628601}, \href
  {https://ui.adsabs.harvard.edu/abs/2016A&A...595A..62P} {595, A62}

\bibitem[\protect\citeauthoryear{{Pettini} \& {Pagel}}{{Pettini} \&
  {Pagel}}{2004}]{Pettini+Pagel04}
{Pettini} M.,  {Pagel} B. E.~J.,  2004, \mn@doi [\mnras]
  {10.1111/j.1365-2966.2004.07591.x}, \href
  {https://ui.adsabs.harvard.edu/abs/2004MNRAS.348L..59P} {348, L59}

\bibitem[\protect\citeauthoryear{{Pilkington} et~al.,}{{Pilkington}
  et~al.}{2012}]{Pilkington+12}
{Pilkington} K.,  et~al., 2012, \mn@doi [\aap] {10.1051/0004-6361/201117466},
  \href {https://ui.adsabs.harvard.edu/abs/2012A&A...540A..56P} {540, A56}

\bibitem[\protect\citeauthoryear{{Pilyugin} \& {Grebel}}{{Pilyugin} \&
  {Grebel}}{2016}]{Pilyugin+Grebel16}
{Pilyugin} L.~S.,  {Grebel} E.~K.,  2016, \mn@doi [\mnras]
  {10.1093/mnras/stw238}, \href
  {https://ui.adsabs.harvard.edu/abs/2016MNRAS.457.3678P} {457, 3678}

\bibitem[\protect\citeauthoryear{{Poetrodjojo} et~al.,}{{Poetrodjojo}
  et~al.}{2021}]{Poetrodjojo+21}
{Poetrodjojo} H.,  et~al., 2021, \mn@doi [\mnras] {10.1093/mnras/stab205},
  \href {https://ui.adsabs.harvard.edu/abs/2021MNRAS.502.3357P} {502, 3357}

\bibitem[\protect\citeauthoryear{{Portinari} \& {Chiosi}}{{Portinari} \&
  {Chiosi}}{2000}]{Portinari+Chiosi2000}
{Portinari} L.,  {Chiosi} C.,  2000, \mn@doi [\aap]
  {10.48550/arXiv.astro-ph/0002145}, \href
  {https://ui.adsabs.harvard.edu/abs/2000A&A...355..929P} {355, 929}

\bibitem[\protect\citeauthoryear{{Queyrel} et~al.,}{{Queyrel}
  et~al.}{2012}]{Queyrel+12}
{Queyrel} J.,  et~al., 2012, \mn@doi [\aap] {10.1051/0004-6361/201117718},
  \href {https://ui.adsabs.harvard.edu/abs/2012A&A...539A..93Q} {539, A93}

\bibitem[\protect\citeauthoryear{{Rosales-Ortega}, {D{\'\i}az}, {Kennicutt}  \&
  {S{\'a}nchez}}{{Rosales-Ortega} et~al.}{2011}]{Rosales-Ortega+11}
{Rosales-Ortega} F.~F.,  {D{\'\i}az} A.~I.,  {Kennicutt} R.~C.,   {S{\'a}nchez}
  S.~F.,  2011, \mn@doi [\mnras] {10.1111/j.1365-2966.2011.18870.x}, \href
  {https://ui.adsabs.harvard.edu/abs/2011MNRAS.415.2439R} {415, 2439}

\bibitem[\protect\citeauthoryear{{Rosolowsky} \& {Simon}}{{Rosolowsky} \&
  {Simon}}{2008}]{Rosolowsky+Simon08}
{Rosolowsky} E.,  {Simon} J.~D.,  2008, \mn@doi [\apj] {10.1086/527407}, \href
  {https://ui.adsabs.harvard.edu/abs/2008ApJ...675.1213R} {675, 1213}

\bibitem[\protect\citeauthoryear{{Rupke}, {Kewley}  \& {Chien}}{{Rupke}
  et~al.}{2010}]{Rupke+10}
{Rupke} D. S.~N.,  {Kewley} L.~J.,   {Chien} L.~H.,  2010, \mn@doi [\apj]
  {10.1088/0004-637X/723/2/1255}, \href
  {https://ui.adsabs.harvard.edu/abs/2010ApJ...723.1255R} {723, 1255}

\bibitem[\protect\citeauthoryear{{Sakhibov}, {Zinchenko}, {Pilyugin}, {Grebel},
  {Just}  \& {V{\'\i}lchez}}{{Sakhibov} et~al.}{2018}]{Sakhibov+18}
{Sakhibov} F.,  {Zinchenko} I.~A.,  {Pilyugin} L.~S.,  {Grebel} E.~K.,  {Just}
  A.,   {V{\'\i}lchez} J.~M.,  2018, \mn@doi [\mnras] {10.1093/mnras/stx2799},
  \href {https://ui.adsabs.harvard.edu/abs/2018MNRAS.474.1657S} {474, 1657}

\bibitem[\protect\citeauthoryear{{S{\'a}nchez-Menguiano}
  et~al.,}{{S{\'a}nchez-Menguiano} et~al.}{2016}]{Sanchez-Menguiano+16}
{S{\'a}nchez-Menguiano} L.,  et~al., 2016, \mn@doi [\apjl]
  {10.3847/2041-8205/830/2/L40}, \href
  {https://ui.adsabs.harvard.edu/abs/2016ApJ...830L..40S} {830, L40}

\bibitem[\protect\citeauthoryear{{S{\'a}nchez-Menguiano}
  et~al.,}{{S{\'a}nchez-Menguiano} et~al.}{2018}]{Sanchez-Menguiano+18}
{S{\'a}nchez-Menguiano} L.,  et~al., 2018, \mn@doi [\aap]
  {10.1051/0004-6361/201731486}, \href
  {https://ui.adsabs.harvard.edu/abs/2018A&A...609A.119S} {609, A119}

\bibitem[\protect\citeauthoryear{{S{\'a}nchez-Menguiano}, {S{\'a}nchez},
  {P{\'e}rez}, {Ruiz-Lara}, {Galbany}, {Anderson}  \&
  {Kuncarayakti}}{{S{\'a}nchez-Menguiano} et~al.}{2020}]{Sanchez-Menguiano+20}
{S{\'a}nchez-Menguiano} L.,  {S{\'a}nchez} S.~F.,  {P{\'e}rez} I.,  {Ruiz-Lara}
  T.,  {Galbany} L.,  {Anderson} J.~P.,   {Kuncarayakti} H.,  2020, \mn@doi
  [\mnras] {10.1093/mnras/staa088}, \href
  {https://ui.adsabs.harvard.edu/abs/2020MNRAS.492.4149S} {492, 4149}

\bibitem[\protect\citeauthoryear{{S{\'a}nchez}, {Rosales-Ortega}, {Kennicutt},
  {Johnson}, {Diaz}, {Pasquali}  \& {Hao}}{{S{\'a}nchez}
  et~al.}{2011}]{Sanchez+11}
{S{\'a}nchez} S.~F.,  {Rosales-Ortega} F.~F.,  {Kennicutt} R.~C.,  {Johnson}
  B.~D.,  {Diaz} A.~I.,  {Pasquali} A.,   {Hao} C.~N.,  2011, \mn@doi [\mnras]
  {10.1111/j.1365-2966.2010.17444.x}, \href
  {https://ui.adsabs.harvard.edu/abs/2011MNRAS.410..313S} {410, 313}

\bibitem[\protect\citeauthoryear{{S{\'a}nchez} et~al.,}{{S{\'a}nchez}
  et~al.}{2012}]{CALIFA}
{S{\'a}nchez} S.~F.,  et~al., 2012, \mn@doi [Astronomy \& Astrohpysics]
  {10.1051/0004-6361/201117353}, \href
  {https://ui.adsabs.harvard.edu/abs/2012A&A...538A...8S} {538, A8}

\bibitem[\protect\citeauthoryear{{S{\'a}nchez} et~al.,}{{S{\'a}nchez}
  et~al.}{2014}]{Sanchez+14}
{S{\'a}nchez} S.~F.,  et~al., 2014, \mn@doi [\aap]
  {10.1051/0004-6361/201322343}, \href
  {https://ui.adsabs.harvard.edu/abs/2014A&A...563A..49S} {563, A49}

\bibitem[\protect\citeauthoryear{{S{\'a}nchez} et~al.,}{{S{\'a}nchez}
  et~al.}{2015}]{Sanchez+15}
{S{\'a}nchez} S.~F.,  et~al., 2015, \mn@doi [\aap]
  {10.1051/0004-6361/201424950}, \href
  {https://ui.adsabs.harvard.edu/abs/2015A&A...573A.105S} {573, A105}

\bibitem[\protect\citeauthoryear{{S{\'a}nchez}, {Walcher}, {Lopez-Cob{\'a}},
  {Barrera-Ballesteros}, {Mej{\'\i}a-Narv{\'a}ez}, {Espinosa-Ponce}  \&
  {Camps-Fari{\~n}a}}{{S{\'a}nchez} et~al.}{2021}]{Sanchez+21}
{S{\'a}nchez} S.~F.,  {Walcher} C.~J.,  {Lopez-Cob{\'a}} C.,
  {Barrera-Ballesteros} J.~K.,  {Mej{\'\i}a-Narv{\'a}ez} A.,  {Espinosa-Ponce}
  C.,   {Camps-Fari{\~n}a} A.,  2021, \mn@doi [\rmxaa]
  {10.22201/ia.01851101p.2021.57.01.01}, \href
  {https://ui.adsabs.harvard.edu/abs/2021RMxAA..57....3S} {57, 3}

\bibitem[\protect\citeauthoryear{{Sanders} et~al.,}{{Sanders}
  et~al.}{2020}]{Sanders+20}
{Sanders} R.~L.,  et~al., 2020, \mn@doi [\mnras] {10.1093/mnras/stz3032}, \href
  {https://ui.adsabs.harvard.edu/abs/2020MNRAS.491.1427S} {491, 1427}

\bibitem[\protect\citeauthoryear{{Sanders} et~al.,}{{Sanders}
  et~al.}{2021}]{Sanders+21}
{Sanders} R.~L.,  et~al., 2021, \mn@doi [\apj] {10.3847/1538-4357/abf4c1},
  \href {https://ui.adsabs.harvard.edu/abs/2021ApJ...914...19S} {914, 19}

\bibitem[\protect\citeauthoryear{{Sanders}, {Shapley}, {Topping}, {Reddy}  \&
  {Brammer}}{{Sanders} et~al.}{2023}]{Sanders+23}
{Sanders} R.~L.,  {Shapley} A.~E.,  {Topping} M.~W.,  {Reddy} N.~A.,
  {Brammer} G.~B.,  2023, \mn@doi [arXiv e-prints] {10.48550/arXiv.2303.08149},
  \href {https://ui.adsabs.harvard.edu/abs/2023arXiv230308149S} {p.
  arXiv:2303.08149}

\bibitem[\protect\citeauthoryear{{Sch{\"o}nrich} \& {McMillan}}{{Sch{\"o}nrich}
  \& {McMillan}}{2017}]{Schonrich+McMillan17}
{Sch{\"o}nrich} R.,  {McMillan} P.~J.,  2017, \mn@doi [\mnras]
  {10.1093/mnras/stx093}, \href
  {https://ui.adsabs.harvard.edu/abs/2017MNRAS.467.1154S} {467, 1154}

\bibitem[\protect\citeauthoryear{{Searle}}{{Searle}}{1971}]{Searle71}
{Searle} L.,  1971, \mn@doi [\apj] {10.1086/151090}, \href
  {https://ui.adsabs.harvard.edu/abs/1971ApJ...168..327S} {168, 327}

\bibitem[\protect\citeauthoryear{{Simons} et~al.,}{{Simons}
  et~al.}{2021}]{Simons+21}
{Simons} R.~C.,  et~al., 2021, \mn@doi [\apj] {10.3847/1538-4357/ac28f4}, \href
  {https://ui.adsabs.harvard.edu/abs/2021ApJ...923..203S} {923, 203}

\bibitem[\protect\citeauthoryear{{Spitoni}, {Cescutti}, {Minchev}, {Matteucci},
  {Silva Aguirre}, {Martig}, {Bono}  \& {Chiappini}}{{Spitoni}
  et~al.}{2019}]{Spitoni+19}
{Spitoni} E.,  {Cescutti} G.,  {Minchev} I.,  {Matteucci} F.,  {Silva Aguirre}
  V.,  {Martig} M.,  {Bono} G.,   {Chiappini} C.,  2019, \mn@doi [\aap]
  {10.1051/0004-6361/201834665}, \href
  {https://ui.adsabs.harvard.edu/abs/2019A&A...628A..38S} {628, A38}

\bibitem[\protect\citeauthoryear{{Stott} et~al.,}{{Stott}
  et~al.}{2014}]{Stott+14}
{Stott} J.~P.,  et~al., 2014, \mn@doi [\mnras] {10.1093/mnras/stu1343}, \href
  {https://ui.adsabs.harvard.edu/abs/2014MNRAS.443.2695S} {443, 2695}

\bibitem[\protect\citeauthoryear{{Swinbank}, {Sobral}, {Smail}, {Geach},
  {Best}, {McCarthy}, {Crain}  \& {Theuns}}{{Swinbank}
  et~al.}{2012}]{Swinbank+12}
{Swinbank} A.~M.,  {Sobral} D.,  {Smail} I.,  {Geach} J.~E.,  {Best} P.~N.,
  {McCarthy} I.~G.,  {Crain} R.~A.,   {Theuns} T.,  2012, \mn@doi [\mnras]
  {10.1111/j.1365-2966.2012.21774.x}, \href
  {https://ui.adsabs.harvard.edu/abs/2012MNRAS.426..935S} {426, 935}

\bibitem[\protect\citeauthoryear{{Tissera}, {Pedrosa}, {Sillero}  \&
  {Vilchez}}{{Tissera} et~al.}{2016}]{Tissera+16}
{Tissera} P.~B.,  {Pedrosa} S.~E.,  {Sillero} E.,   {Vilchez} J.~M.,  2016,
  \mn@doi [\mnras] {10.1093/mnras/stv2736}, \href
  {https://ui.adsabs.harvard.edu/abs/2016MNRAS.456.2982T} {456, 2982}

\bibitem[\protect\citeauthoryear{{Tissera}, {Rosas-Guevara}, {Sillero},
  {Pedrosa}, {Theuns}  \& {Bignone}}{{Tissera} et~al.}{2022}]{Tissera+22}
{Tissera} P.~B.,  {Rosas-Guevara} Y.,  {Sillero} E.,  {Pedrosa} S.~E.,
  {Theuns} T.,   {Bignone} L.,  2022, \mn@doi [\mnras]
  {10.1093/mnras/stab3644}, \href
  {https://ui.adsabs.harvard.edu/abs/2022MNRAS.511.1667T} {511, 1667}

\bibitem[\protect\citeauthoryear{{Torrey}, {Cox}, {Kewley}  \&
  {Hernquist}}{{Torrey} et~al.}{2012}]{Torrey+12}
{Torrey} P.,  {Cox} T.~J.,  {Kewley} L.,   {Hernquist} L.,  2012, \mn@doi
  [\apj] {10.1088/0004-637X/746/1/108}, \href
  {https://ui.adsabs.harvard.edu/abs/2012ApJ...746..108T} {746, 108}

\bibitem[\protect\citeauthoryear{{Tremonti} et~al.,}{{Tremonti}
  et~al.}{2004}]{Tremonti+04}
{Tremonti} C.~A.,  et~al., 2004, \mn@doi [\apj] {10.1086/423264}, \href
  {https://ui.adsabs.harvard.edu/abs/2004ApJ...613..898T} {613, 898}

\bibitem[\protect\citeauthoryear{{Troncoso} et~al.,}{{Troncoso}
  et~al.}{2014}]{Troncoso+14}
{Troncoso} P.,  et~al., 2014, \mn@doi [\aap] {10.1051/0004-6361/201322099},
  \href {https://ui.adsabs.harvard.edu/abs/2014A&A...563A..58T} {563, A58}

\bibitem[\protect\citeauthoryear{{Vila-Costas} \& {Edmunds}}{{Vila-Costas} \&
  {Edmunds}}{1992}]{VilaCostas+92}
{Vila-Costas} M.~B.,  {Edmunds} M.~G.,  1992, \mn@doi [\mnras]
  {10.1093/mnras/259.1.121}, \href
  {https://ui.adsabs.harvard.edu/abs/1992MNRAS.259..121V} {259, 121}

\bibitem[\protect\citeauthoryear{{Vogt}, {P{\'e}rez}, {Dopita},
  {Verdes-Montenegro}  \& {Borthakur}}{{Vogt} et~al.}{2017}]{Vogt+17}
{Vogt} F.~P.~A.,  {P{\'e}rez} E.,  {Dopita} M.~A.,  {Verdes-Montenegro} L.,
  {Borthakur} S.,  2017, \mn@doi [\aap] {10.1051/0004-6361/201629853}, \href
  {https://ui.adsabs.harvard.edu/abs/2017A&A...601A..61V} {601, A61}

\bibitem[\protect\citeauthoryear{{Wang} et~al.,}{{Wang} et~al.}{2019}]{Wang+19}
{Wang} X.,  et~al., 2019, \mn@doi [\apj] {10.3847/1538-4357/ab3861}, \href
  {https://ui.adsabs.harvard.edu/abs/2019ApJ...882...94W} {882, 94}

\bibitem[\protect\citeauthoryear{{Wang} et~al.,}{{Wang} et~al.}{2020}]{Wang+20}
{Wang} X.,  et~al., 2020, \mn@doi [\apj] {10.3847/1538-4357/abacce}, \href
  {https://ui.adsabs.harvard.edu/abs/2020ApJ...900..183W} {900, 183}

\bibitem[\protect\citeauthoryear{{Wang} et~al.,}{{Wang} et~al.}{2022}]{Wang+22}
{Wang} X.,  et~al., 2022, \mn@doi [\apjl] {10.3847/2041-8213/ac959e}, \href
  {https://ui.adsabs.harvard.edu/abs/2022ApJ...938L..16W} {938, L16}

\bibitem[\protect\citeauthoryear{{Worthey}, {Espa{\~n}a}, {MacArthur}  \&
  {Courteau}}{{Worthey} et~al.}{2005}]{Worthey+05}
{Worthey} G.,  {Espa{\~n}a} A.,  {MacArthur} L.~A.,   {Courteau} S.,  2005,
  \mn@doi [\apj] {10.1086/432785}, \href
  {https://ui.adsabs.harvard.edu/abs/2005ApJ...631..820W} {631, 820}

\bibitem[\protect\citeauthoryear{{Wuyts} et~al.,}{{Wuyts}
  et~al.}{2016}]{Wuyts+16}
{Wuyts} E.,  et~al., 2016, \mn@doi [\apj] {10.3847/0004-637X/827/1/74}, \href
  {https://ui.adsabs.harvard.edu/abs/2016ApJ...827...74W} {827, 74}

\bibitem[\protect\citeauthoryear{{Yong}, {Carney}  \& {Friel}}{{Yong}
  et~al.}{2012}]{Yong+12}
{Yong} D.,  {Carney} B.~W.,   {Friel} E.~D.,  2012, \mn@doi [\aj]
  {10.1088/0004-6256/144/4/95}, \href
  {https://ui.adsabs.harvard.edu/abs/2012AJ....144...95Y} {144, 95}

\bibitem[\protect\citeauthoryear{{Yuan}, {Kewley}  \& {Rich}}{{Yuan}
  et~al.}{2013}]{Yuan+13}
{Yuan} T.~T.,  {Kewley} L.~J.,   {Rich} J.,  2013, \mn@doi [\apj]
  {10.1088/0004-637X/767/2/106}, \href
  {https://ui.adsabs.harvard.edu/abs/2013ApJ...767..106Y} {767, 106}

\bibitem[\protect\citeauthoryear{{Zahid}, {Dima}, {Kewley}, {Erb}  \&
  {Dav{\'e}}}{{Zahid} et~al.}{2012}]{Zahid+12}
{Zahid} H.~J.,  {Dima} G.~I.,  {Kewley} L.~J.,  {Erb} D.~K.,   {Dav{\'e}} R.,
  2012, \mn@doi [\apj] {10.1088/0004-637X/757/1/54}, \href
  {https://ui.adsabs.harvard.edu/abs/2012ApJ...757...54Z} {757, 54}

\bibitem[\protect\citeauthoryear{{Zahid}, {Dima}, {Kudritzki}, {Kewley},
  {Geller}, {Hwang}, {Silverman}  \& {Kashino}}{{Zahid}
  et~al.}{2014}]{Zahid+14}
{Zahid} H.~J.,  {Dima} G.~I.,  {Kudritzki} R.-P.,  {Kewley} L.~J.,  {Geller}
  M.~J.,  {Hwang} H.~S.,  {Silverman} J.~D.,   {Kashino} D.,  2014, \mn@doi
  [\apj] {10.1088/0004-637X/791/2/130}, \href
  {https://ui.adsabs.harvard.edu/abs/2014ApJ...791..130Z} {791, 130}

\bibitem[\protect\citeauthoryear{{Zinchenko}, {Pilyugin}, {Grebel},
  {S{\'a}nchez}  \& {V{\'\i}lchez}}{{Zinchenko} et~al.}{2016}]{Zinchenko+16}
{Zinchenko} I.~A.,  {Pilyugin} L.~S.,  {Grebel} E.~K.,  {S{\'a}nchez} S.~F.,
  {V{\'\i}lchez} J.~M.,  2016, \mn@doi [\mnras] {10.1093/mnras/stw1857}, \href
  {https://ui.adsabs.harvard.edu/abs/2016MNRAS.462.2715Z} {462, 2715}

\bibitem[\protect\citeauthoryear{{de Graaff} et~al.,}{{de Graaff}
  et~al.}{2023}]{Graaff+23}
{de Graaff} A.,  et~al., 2023, \mn@doi [arXiv e-prints]
  {10.48550/arXiv.2308.09742}, \href
  {https://ui.adsabs.harvard.edu/abs/2023arXiv230809742D} {p. arXiv:2308.09742}

\bibitem[\protect\citeauthoryear{{van der Wel} et~al.,}{{van der Wel}
  et~al.}{2024}]{vanderWel+24}
{van der Wel} A.,  et~al., 2024, \mn@doi [\apj] {10.3847/1538-4357/ad02ee},
  \href {https://ui.adsabs.harvard.edu/abs/2024ApJ...960...53V} {960, 53}

\makeatother
\end{thebibliography}




\appendix
\section{Modelling heteroskedastic noise in metallicity}
\label{ap:heteronoise}

In this Appendix, we present our derivation for a physically-motivated model of the heteroskedastic noise in metallicity measurements derived using the strong emission line method for many spaxels in a galaxy.

We model the uncertainty in the metallicity measured for each spaxel as coming from two sources. Firstly, we expect some \textit{model error} (intrinsic scatter) due to the imperfect nature of strong emission line diagnostics \citep[e.g.][]{Pilyugin+Grebel16, Curti+20, Sanders+23}. This model error adds some uncertainty to the recovered metallicity of all spaxels, regardless of their S/N. Additionally, we expect some \textit{measurement error} due to the finite signal to noise at each spaxel. We expect this component to depend on the surface brightness $I$ of each spaxel. Since these two sources of error are independent, they will add in quadrature to give the uncertainty in the metallicity of each spaxel:

\begin{equation}
    \sigma_Z^2 = \underbrace{c_1}_{\text{model error}} + \underbrace{c_2 f(I)}_{\text{measurement error}}.
    \label{eq:B1}
\end{equation}
We devote the remainder of this Appendix to finding an appropriate functional form for $f(I)$. That is, we want to know how the uncertainty in the measured metallicity in a spaxel decreases as the brightness of that spaxel is increased.

Mathematically, a strong emission line diagnostic is a function $f_{\text{SEL}}$ that takes as input an \textit{abundance indicator}, and returns the metallicity. An abundance indicator is generally a linear combination of the logarithm of the intensity of some number of emission lines, such as N$_2$O$_2 := \log_{10} ( $[N \textsc{ii}]$\lambda6583/$[O \textsc{ii}]$\lambda\lambda3726,29)$, or O$_3$N$_2 :=\log_{10}( $[O \textsc{iii}]$\lambda5007/$H$\alpha ) - \log_{10}( $[N \textsc{ii}]$\lambda6583/$H$\beta )$. We use the variable $A$ to represent a generic abundance indicator with $n_A$ emission lines $l_1, ..., l_{n_A}$, and define it mathematically as:

\begin{equation}
    A = \sum_{i=1}^{n_A} c_i \ln (l_i) ,
\end{equation}
where $l_i$ is the intensity of the $i$th emission line.

Then, we may define our strong emission line diagnostic as a function $f_{\text{SEL}}$ such that 

\begin{equation}
    Z \approx f_{\text{SEL}}(A),
\end{equation}
plus some model error. All measurement error, then, comes from uncertainties in the measurements intensities of the emission lines $l_i$, which we denote as $\sigma_i$. We may find a formula for how the uncertainties in the measured values of $l_i$ lead to measurement uncertainty in $Z$ using linear error propagation:

\begin{eqnarray}
    \sigma_Z^2 &=& {\left(\frac{df_{\text{SEL}}}{dA}\right)}^2 \sum_{i=1}^{n_A} {\left(\frac{\partial A}{\partial l_i}\right)}^2 \sigma_i^2 \\
    &=& {\left(\frac{df_{\text{SEL}}}{dA}\right)}^2 \sum_{i=1}^{n_A} c_i^2 {\left(\frac{\sigma_i}{l_i}\right)}^2 
\end{eqnarray}
The intensity of each line $l_i$ is proportional to the surface brightness $I$ at each spaxel. If we assume that Poission noise is the dominating source of uncertainty in the brightness of emission lines, then $\sigma_i^2 \propto I$ also for each line. Making these substitutions, and absorbing all terms that do not depend on brightness into a proportionality factor, this leaves us with an equation for how measurement uncertainty depends on $I$:

\begin{equation}
    \sigma_Z^2 \propto \frac{I}{I^2} = I^{-1}.
\end{equation}
Incorporating this into \ref{eq:B1}, we get the final expression for our heteroskedastic noise in the measured metallicity:

\begin{equation}
    \sigma_Z^2 = c_1 + \frac{c_2}{I}.
\end{equation}

\bsp	
\label{lastpage}
\end{document}